\documentclass{emulateapj}
\newcommand{\HI}{\ion{H}{1}}
\newcommand{\HII}{\ion{H}{2}}
\slugcomment{Accepted for publication in ApJ}
\shorttitle{HI CLOUDS IN THE LOWER HALO}
\shortauthors{Ford et al.}
\begin{document}

\title{\HI\ CLOUDS IN THE LOWER HALO: I. THE GALACTIC ALL-SKY SURVEY
  PILOT REGION} 
\author{H. Alyson Ford\altaffilmark{1,2},
  N. M. McClure-Griffiths\altaffilmark{2},
  Felix J. Lockman\altaffilmark{3},
  J. Bailin\altaffilmark{4},
  M. R. Calabretta\altaffilmark{2},
  P. M. W. Kalberla\altaffilmark{5},
  T. Murphy\altaffilmark{6,7},
  D. J. Pisano\altaffilmark{3}}
\altaffiltext{1}{Center for Astrophysics and Supercomputing, Swinburne
University of Technology, Mail H39, P.O. Box 218, Hawthorn, Victoria
3122, Australia; alyson@astro.swin.edu.au}
\altaffiltext{2}{Australia Telescope National Facility, CSIRO,
  P.O. Box 76, Epping, NSW 1710, Australia.}
\altaffiltext{3}{National Radio Astronomy Observatory, P.O. Box 2,
  Green Bank, WV 24944, USA.}
\altaffiltext{4}{Department of Physics \& Astronomy, McMaster
  University, 1280 Main Street West, Hamilton, ON, L8S 4M1, Canada.}
\altaffiltext{5}{Radioastronomisches Institut der Universit\"at Bonn,
Auf dem H\"ugel 71, 53121 Bonn, Germany.}
\altaffiltext{6}{School of Physics, University of Sydney, NSW 2006, Australia.}
\altaffiltext{7}{School of Information Technologies, University of Sydney, NSW 2006, Australia.}

\begin{abstract}

We have detected over $400$ \HI\ clouds in the lower halo of the
Galaxy within the  pilot region of the Galactic All-Sky Survey (GASS),
a region of the fourth  quadrant that spans $18\degr$ in longitude,
$40\degr$ in latitude and is centered on the Galactic equator. These
clouds have a median peak brightness temperature of $0.6$~K, a median
velocity width of $12.8$~km~s$^{-1}$, and angular sizes
$\lesssim 1\degr$. The motion of these clouds is dominated by Galactic
rotation with a random cloud-to-cloud velocity dispersion of
$18$~km~s$^{-1}$. A sample of clouds likely to be near tangent
points was analyzed in detail. These clouds have radii on the order of
$30$~pc and a median \HI\ mass of $630 M_{\odot}$. The population has a 
vertical scale height of $400$~pc and is concentrated in
Galactocentric radius, peaking at $R=3.8$~kpc. This confined structure
suggests that the clouds are linked to spiral features, while
morphological evidence that many clouds are aligned with loops and
filaments is suggestive of a relationship with star formation. The
clouds might result from supernovae and stellar winds in the form of
fragmenting shells and gas that has been pushed into the halo rather
than from a galactic fountain.
\end{abstract}


\section{INTRODUCTION}

Neutral atomic hydrogen (\HI) is ubiquitous throughout the Galaxy with
a wide variety of morphologies and kinematics, and exhibits many complex
structures, including worms \citep{1992Koo}, sheets and filaments
\citep{1967Heiles,1990Dickey}, shells
\citep{1979Heiles,2002McClure-Griffiths}, and clouds
\citep{2002Lockman}. The Galactic \HI\ disk extends to Galactocentric
radii, $R\geq 30$~kpc and its thickness varies from $\leq 100$~pc
inside $R=3.5$ kpc to $\sim 3$ kpc in the outer Galaxy, with a roughly
uniform thickness of $230$~pc between $3.5$~kpc and the solar circle
(see \citealt{2001Ferriere} and references within). \HI\ is also known
to extend far beyond the thin \HI\ disk as a layer into the Galactic
halo \citep{1984Lockman}. However, recent high angular resolution
observations using the Green Bank Telescope (GBT) have revealed that this
layer is not smooth but instead is composed of small \HI\ clouds with
sizes on the order of a few tens of parsecs and masses of $50
M_\odot$ \citep{2002Lockman}; confusion may limit the detectability of
such clouds at low heights, in which case they may not be confined to
the halo (see, e.g., \citealt{2006Stil}). These clouds follow
Galactic rotation and are discrete clumps of \HI\ that are localized
in space and velocity. Although they are sometimes related to larger
structures, for example, embedded in filaments, each cloud appears to
be a distinct object. The gross properties of the Milky Way's thick
\HI\ layer may in fact be a consequence of the statistical properties
of these \HI\ clouds and it is possible that thick \HI\ disks in other
galaxies, once observed with sufficient resolution, would reveal a
similar structure.         

While the origin of the halo clouds is unknown, one possible
explanation is a ``galactic fountain'' model, where hot gas produced
by supernovae rises into the halo of the Galaxy, cools and condenses
into \HI\ clouds, which then fall back to the plane
\citep{1976Shapiro,1980Bregman}. \citet{1990Houck} predicted that,
from a lower temperature fountain, gas in Galactic rotation could be
formed at heights close to the plane. This scenario is supported by
observations of intermediate velocity clouds (IVCs), such as those of
cloud g1, whose location, kinematics and abundances match those
expected \citep{2008Wakker}. The abundances in the
IV Arch \citep{2001RichterA} and LLIV Arch \citep{2001RichterB} also
suggest a fountain origin; as their abundances are near solar it
is likely that they originated from material enriched from the disk. The
location of these IVCs also support this, as they are roughly
$1$~kpc from the disk. 

Another possibility is that the halo clouds originate in environments
where supernovae and stellar winds disrupt the surrounding
medium. Such events can result in the formation of a bubble, and models
suggest that with a large enough energy source it is possible for
these bubbles to expand beyond the thickness of the Galactic disk
(e.g., \citealt{1988Tomisaka,1990Heiles}). These bubbles are
encompassed by an \HI\ shell as a result of either radiative cooling,
which then accumulates more \HI\ as the bubble continues to expand, or
solely the sweeping up of ambient material, depending on the wind
speed \citep{1992KooB}. The shell remains as a single entity until in
some cases Rayleigh-Taylor instabilities cause it to fragment
\citep{1989MacLow}. Once this fragmentation has occurred, gas that has
been shock heated is expelled outwards, mixing material from the disk
with that in the halo \citep{1989Norman} and the remaining fragments
of the \HI\ shell may be the observed halo clouds
\citep{2006McClure-Griffiths}. It is also possible for the hot gas
that has been expelled to cool and recombine, as seen in models of
\citet{2000deAvillez}, or perhaps energy from increased supernova
activity in areas of active star formation has simply pushed disk gas
into the halo, forming clumps of \HI.

Large samples of halo clouds are required to constrain the properties
and distribution of the population and provide insight into
their origin. As the \citet{2002Lockman} sample only has $38$ clouds
in a small region of the Galaxy, the properties of the
population that constitute the \HI\ layer are currently not
well-determined. In this paper we present a catalog of over $400$ \HI\
clouds in the lower halo of the inner Galaxy, which we have detected
in the Galactic All-Sky Survey pilot region, along with an analysis
and discussion of their properties and distribution. We begin with an
overview of the observations and data in \S\ref{observations} and
present the observed properties of all clouds in \S \ref{clouds}. In
\S \ref{tangentpoints} we determine the physical properties of a
subset of these clouds that can be assumed to be located at tangent
points, where the observer's line-of-sight is tangent to circles of
constant Galactocentric radius. An analysis of the cloud properties is
presented in \S \ref{discussion} and implications of these results are
discussed in \S \ref{nature}. We summarize the results and discuss
future work in \S\ref{conclusions}.          

\section{OBSERVATIONS AND DATA}
\label{observations}

The data presented in this paper are from the Galactic All-Sky Survey
(GASS), a fully sampled Galactic \HI\ survey that covers the entire
sky south of declination $\delta=0\degr$. GASS data were taken with
the 21 cm Multibeam receiver \citep{1996StaveleySmith} at the Parkes
Radio Telescope and cover $-400 \le V_{\mathrm{LSR}} \le
450$~km~s$^{-1}$, where $V_{\mathrm{LSR}}$ is the velocity with
respect to the local standard of rest. The  spectral resolution of the
data is $0.8$~km~s$^{-1}$ and the half power beam width of the
Multibeam is $\sim 15\arcmin$. 

GASS observations began in 2005 January and were completed by the end
of 2006. Data reduction was performed using the Livedata package,
which is part of the ATNF subset of the aips++ distribution. The
bandpass correction was performed using an algorithm designed
specifically for GASS data and the Doppler correction was
applied. Fluxes were calibrated from observations of the standards S6,
S8 and S9 \citep{1973Williams}. The reduced data were gridded into a
3D data cube with voxel dimensions of $4'\times 4'\times0.8$ km
s$^{-1}$ using the Gridzilla package, which is also part of the ATNF
subset of the aips++ distribution, and was based on the gridding
algorithm described in \citet{2001Barnes}. The rms spectral noise,
$\Delta T_b \sim 60$ mK, was determined by measuring brightness
temperature fluctuations in a $\sim 200$ square degree region of the
survey that spanned 48 velocity channels free of any obvious
emission. The final GASS data release will be corrected for stray
radiation according to the procedure described in
\citet{2005Kalberla}. Further details on GASS can be found in
\citet{2006McClure-Griffiths}, while extensive details will be
presented in a future paper.

\begin{figure}
\plotone{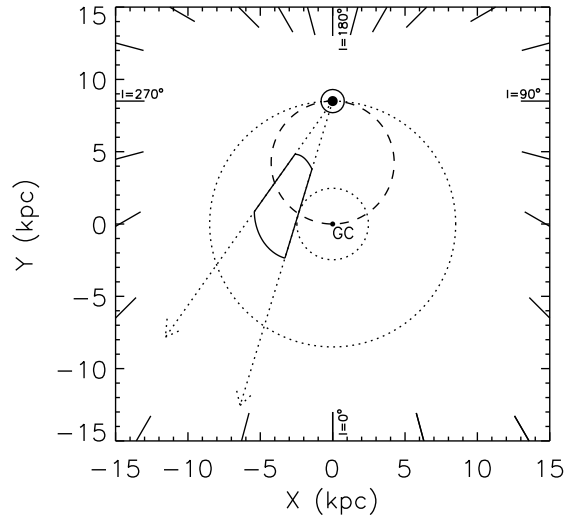}
\caption{\label{tangptdiagram} The area of the Galaxy discussed in
  this paper is bound by the longitude limits of the GASS pilot region
  (dotted arrows) and by $V_{\mathrm{LSR}}= -70$ km~s$^{-1}$, which,
  for a flat rotation curve, gives the area enclosed by the solid
  line. Dotted circles mark Galactocentric radii of $2.47$~kpc and
  $8.5$~kpc and the locus of tangent points is shown by the dashed
  curve that connects the Sun and the Galactic center.}
\end{figure}

In this paper we present results from the GASS pilot region, a region
that was chosen for a preliminary study on halo clouds to develop
techniques that will be applied to the entire survey in the
future. The GASS pilot region is in the fourth quadrant in the inner
Galaxy and spans $325\degr \lesssim l \lesssim 343\degr$, $|b|\lesssim
20\degr$, and $-200 \le V_{\mathrm{LSR}} \le -70$ km~s$^{-1}$ 
(see Figure~\ref{tangptdiagram}). Our focus is on
spatially discrete \HI\ features with an angular size $\lesssim
1\degr$. These features are not characteristic of stray radiation, so
although these data have not been corrected for stray radiation, it is
unlikely that any of the clouds discussed here are a spurious result
of this effect. 

\section{HI CLOUDS IN THE GASS PILOT REGION}
\label{clouds}

\begin{figure}
\plotone{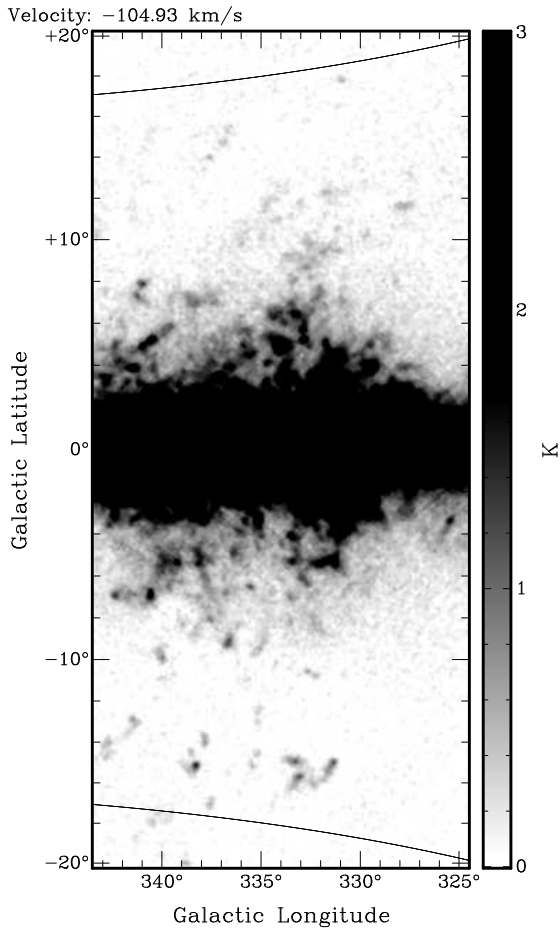}
\caption{\label{primage} GASS pilot region at
  $V_{\mathrm{LSR}}=-105$~km~s$^{-1}$. Many \HI\ clouds with angular
  sizes $\lesssim 1\degr$ are observed,
  both near the Galactic plane and into the lower halo. 
  The curved lines at the top and
  bottom are the boundaries of the region searched.}
\end{figure}

\begin{figure}
\plotone{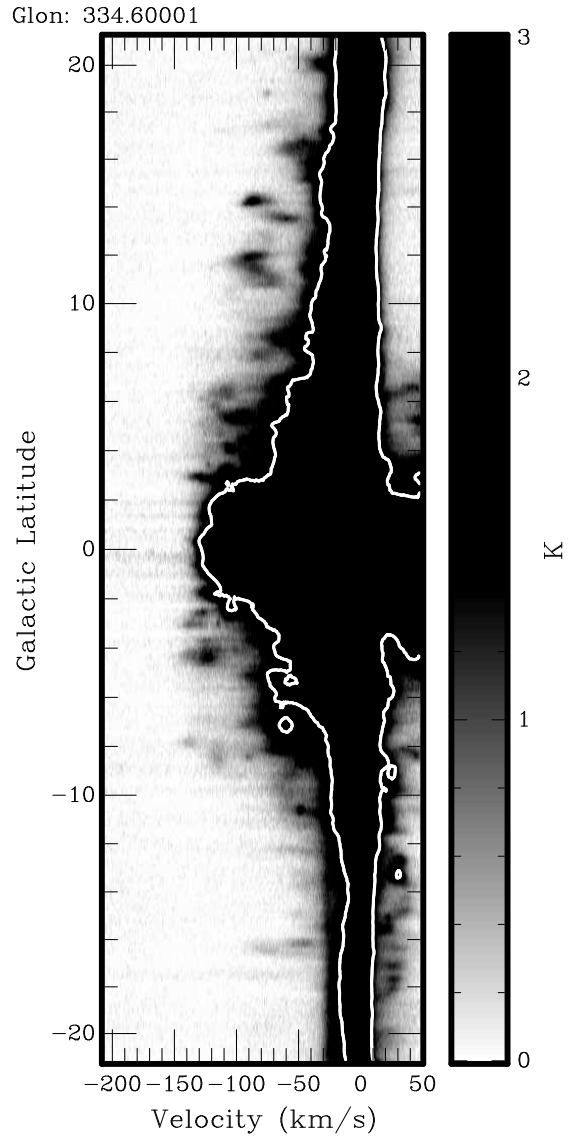}
\caption{\label{primage2} \HI\
  clouds are observed at a variety of velocities, as can be seen in this
  latitude--velocity diagram at $l=334.60\degr$. The contour represents
  $T_b=3$~K, which is the brightness temperature above which
  few clouds can be distinguished due to confusion.}
\end{figure}

We detect numerous discrete \HI\ features in the lower halo of the
Galaxy with angular sizes $\lesssim 1\degr$, similar to the population
of halo clouds discovered by \citet{2002Lockman}. Samples of these clouds can be seen in Figures
\ref{primage} and \ref{primage2}, where we display a longitude-latitude
image at $V_{\mathrm{LSR}}=-105$~km~s$^{-1}$ and a latitude-velocity
image at $l=334.60\degr$, respectively.
The curved lines at the top and bottom of Figure \ref{primage} represent the
boundaries of the region studied in this paper. The contour in Figure
\ref{primage2} represents $T_b=3$ K. Where the emission is brighter than this it
is difficult to distinguish clouds because of confusion, although we
do detect some clouds above this threshold. These figures clearly
demonstrate the presence of discrete \HI\ clouds at a variety of
longitudes, latitudes and velocities, which are seen both close to the
disk and into the halo. Some clouds are extended and some compact, and many
appear to be related to diffuse and filamentary structures. 

\subsection{Search Method and Criteria}
\label{pilotregion}

We chose the following selection criteria to generate a homogeneously
selected catalog of halo clouds:\\
1. The cloud must be within the GASS pilot region, i.e., within
$324.7\degr \le l \le 343.1\degr$, $|b|\lesssim 20\degr$ (see Figure
\ref{primage} for exact boundaries), and $-200 \le V_{\mathrm{LSR}} \le
-70$ km~s$^{-1}$.\\ 
2. The cloud must span four or more pixels and be clearly visible over
three or more channels in the spectra. Most cloud detections were
made with $T_b\geq 5\Delta T_{b}$, where $\Delta T_{b}=60$~mK.\\
3. The cloud must be distinguishable from unrelated background
emission. It was impossible to separate clouds from this emission at
low latitudes in the least negative velocity channels, where the
emission is particularly complex.  

We believe that we have identified all obvious clouds that meet the
criteria listed above. However, it should be noted that some clouds
appeared to have double-peaked velocity profiles. In these
circumstances, each peak was cataloged as an individual cloud because
confusion may be important. If so, confusion may have resulted in the
merging of profiles of multiple clouds at similar locations but with
distinct velocities, resulting in a double-peaked profile. We will
discuss such effects in detail in a subsequent publication.  

Most of the clouds are not isolated and spherical, but are instead
often part of nebulous, filamentary structures and/or sitting in a
fluctuating diffuse background. As a result, all automated cloud
finding algorithms that we tested had difficulty differentiating
clouds from neighbouring clouds and from background emission. Clouds
were therefore identified and their properties measured by visual
inspection of the data cubes.   

\subsection{Properties of the Entire Cloud Sample}
\label{observedcolumns}

\begin{figure}
\plotone{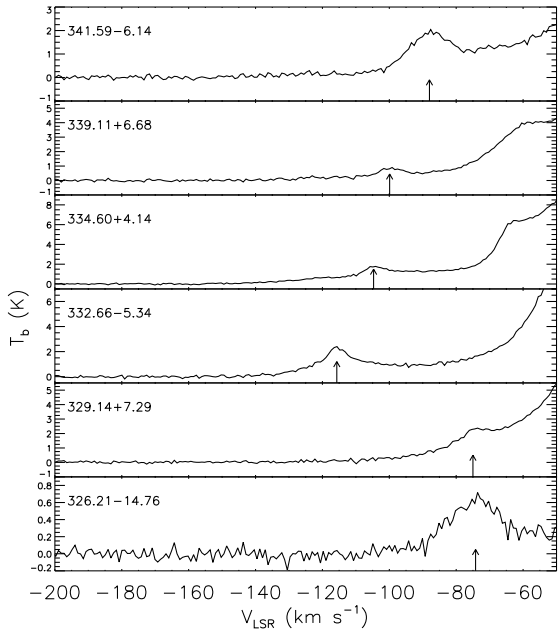}
\caption{\label{spectra} Spectra from a random sample of \HI\ clouds
  from Table \ref{catalog}. Many clouds appear to be
  sitting on broad spectral wings. Arrows mark the
  velocity where the brightness temperature peaks (after background
  subtraction) for each profile.}
\end{figure}

\begin{figure}
\plotone{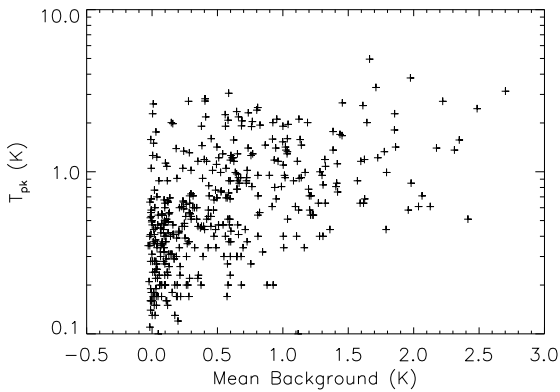}
\caption{\label{background} Peak brightness temperature as a function
  of the mean background subtracted from the integrated intensity
  map. In regions of high background levels there is a clear decline in
  the number of clouds with low peak brightness temperatures.}
\end{figure}

Using the criteria presented in \S \ref{pilotregion} we measured $403$
\HI\ clouds in the GASS pilot region. The observed properties of these
clouds are presented in Table \ref{catalog}. We provide a description
of each property and how it was determined below, along with sample
spectra in Figure \ref{spectra}. An integrated intensity map, which
has the summed intensities over the velocity range for a given cloud,
was used to aid in the determination of some properties. These maps
have had a background subtracted that was the mean flux of unrelated
emission in three interactively chosen areas surrounding the
cloud. It is apparent from Figure
\ref{background} that at the low end of the peak brightness temperature
distribution ($\lesssim 0.5$~K) there is significant incompleteness in
regions of higher background levels. We are confident in our
background subtraction because beyond incompleteness there is no
trend between the peak brightness temperature and mean background
level. 

The columns in Table \ref{catalog} are:

Columns 1-2: The Galactic longitude, $l$, and latitude, $b$, of the
cloud at the position of the peak brightness temperature.

Column 3: The cloud velocity with respect to the local standard of
rest, $V_{\mathrm{LSR}}$, in km~s$^{-1}$, measured as the velocity of the
cloud's peak brightness temperature after background subtraction. The
background level was determined by fitting a line between each edge of
the cloud velocity profile where it merges with surrounding emission.  

Column 4: The cloud's peak brightness temperature after background subtraction,
$T_{\mathrm{pk}}$, in K. Most clouds that we measured had $T_b < 3$~K
prior to background subtraction, as clouds with $T_b$ greater than
this tended to be found only in areas of high confusion.     

Column 5: $\Delta v$ is the full-width at half-maximum (FWHM) of the
velocity profile, determined by inspection after background
subtraction, in km~s$^{-1}$.

Column 6: The \HI\ column density, $N_{HI}$, at the cloud center is
$1.94\times 10^{18}T_{\mathrm{pk}}\Delta v$~cm$^{-2}$ in the optically
thin limit, an assumption that is reasonable because the emission is
faint. $N_{HI}$ was determined after we subtracted a background from
the integrated intensity map. The background level was highly variable
from cloud to cloud being mainly dependent on the latitude and velocity of
the cloud.

Column 7: $\theta_{\mathrm{maj}}$ and $\theta_{\mathrm{min}}$
are the maximum and minimum extent of the cloud in arcminutes and were
determined by inspection from the integrated intensity map of the
cloud. Many of the clouds are unresolved in at least one dimension but
we have not deconvolved their angular sizes due to the uncertainties
associated with the fluctuating background levels of the integrated
intensity maps. These values therefore represent upper limits of the
angular extent.

Column 8: $M_{HI}d^{-2}$ represents the \HI\ mass of the cloud in
units of $M_\odot$ kpc$^{-2}$, where $d$ is the distance to the cloud
in kpc. A background was subtracted from the integrated intensity map,
thereby leaving only the flux of the cloud, which was then summed.  
The mass also relies on the assumption that the \HI\ is optically thin.

Column 9: Details of prior detections in the literature, if any, are
noted.

\subsection{Uncertainties in Observed Properties}
\label{observeduncertainties}

$\delta V_{\mathrm{LSR}}$: The difference between $V_{\mathrm{LSR}}$
and the velocity where the profile decreases from the peak by  $\sqrt{\Delta
T_b^2+\delta T_{\mathrm{pk}}^2}$, assuming the profile can be
approximated by a Gaussian, where $\delta T_{\mathrm{pk}}$ is the
error on $T_{\mathrm{pk}}$ (see below) and $\Delta T_b$ is the rms
noise. 

$\delta T_{\mathrm{pk}}$: This error is assumed to be 
$\sqrt{\Delta T_b^2+(\Delta T_b/\sqrt{2})^2}$, where the first term
represents the error in $T_{\mathrm{pk}}$ prior to the background
subtraction and the second term represents the error of the mean of
the two points defining the background.  

$\delta \Delta v$: Each of the two half maximum points on the profile
is defined as the point where the profile, with error $\Delta T_b$,
crosses $(1/2)T_{\mathrm{pk}}$, with error $(1/2)\delta
T_{\mathrm{pk}}$. We translate these $T_b$ errors into velocity errors
using the slope of a Gaussian profile. We also add the channel width
in quadrature. 

$\delta N_{HI}$: The error on the \HI\ column density, $\delta
N_{HI}=N_{HI}\sqrt{(\delta T_{\mathrm{pk}}/T_{\mathrm{pk}})^2+(\delta
  \Delta v/\Delta v)^2}$.  

$\delta \theta_{\mathrm{maj}}$ and $\delta \theta_{\mathrm{min}}$: The
uncertainties due to the interactive nature of the angular extent
measurements and the fluctuating background levels of the integrated
intensity maps dominate over the nominal statistical error. The
uncertainties were therefore calibrated by comparing values obtained
during different trials for a randomly selected subset of clouds and
are estimated to be approximately $25\%$ of the determined angular
extent.       

$\delta M_{HI}d^{-2}$: The errors introduced by the interactive
process of determining the cloud area and subtracting the background
dominate the mass uncertainties and are estimated to be $40\%$ of the
determined mass, based on the comparison of different trials of
randomly selected clouds mentioned above. 

\subsection{Observed Properties}
\label{obsprops}

\begin{figure}
\plotone{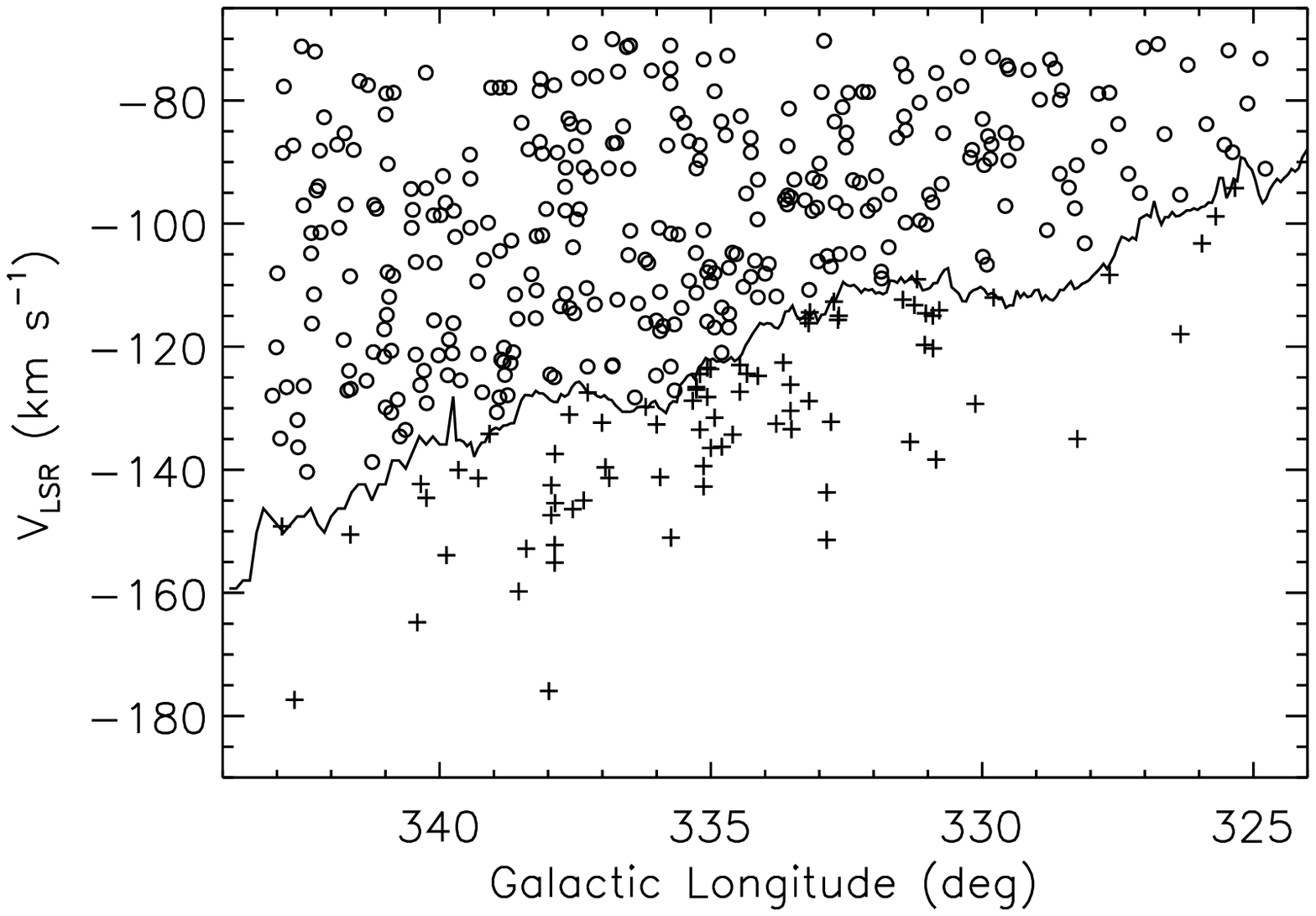}
\caption{\label{lv}$V_{\mathrm{LSR}}$ as a function of longitude for
  all clouds in Table~\ref{catalog}. Crosses represent a subset of
  clouds with $V_{\mathrm{dev}} \leq 0.8$ km~s$^{-1}$ (the tangent
  point sample; \S \ref{tangentpoints}) while circles represent
  all other GASS pilot region clouds. The solid line is the
  $V_{\mathrm{t}}$ curve determined by \citet{2007McClure-Griffiths}
  from \HI\ observations for $l \leq 339.695\degr$ and by
  \citet{2006Luna} from CO observations for all remaining
  longitudes. As we have searched for all clouds within $-200 \leq
  V_{\mathrm{LSR}} \leq -70$ km~s$^{-1}$, the decline in the number of
  clouds at velocities more negative than the terminal velocity
  demonstrates that the kinematics of the clouds are dominated by
  Galactic rotation.}   
\end{figure}

The $V_{\mathrm{LSR}}$ as a function of longitude for all clouds is
shown in Figure \ref{lv}, along with a solid line denoting the
terminal velocity, $V_{\mathrm{t}}$, which, in the fourth quadrant, is
the most negative velocity permitted by Galactic rotation.  The clouds
are abundant at velocities allowed by Galactic rotation and there is
clearly a decline in the number of clouds beyond (more negative than)
$V_{\mathrm{t}}$, which demonstrates that the motions of this cloud
population are dominated by Galactic rotation. Clouds beyond
$V_{\mathrm{t}}$ are said to have ``forbidden velocities'' and the amount
their $V_{\mathrm{LSR}}$ differs from $V_{\mathrm{t}}$ is the
deviation velocity,
$V_{\mathrm{dev}}=V_{\mathrm{LSR}}-V_{\mathrm{t}}$, as defined by
\citet{1991bWakker}. For $l \leq 339.695$, we used $V_{\mathrm{t}}$
determined by \citet{2007McClure-Griffiths} from \HI\ observations
from the Southern Galactic Plane Survey
\citep{2005McClure-Griffiths}. For all remaining longitudes we used
$V_{\mathrm{t}}$ determined by \citet{2006Luna} from the
Columbia-Universidad de Chile CO surveys.  

\begin{figure}
\plotone{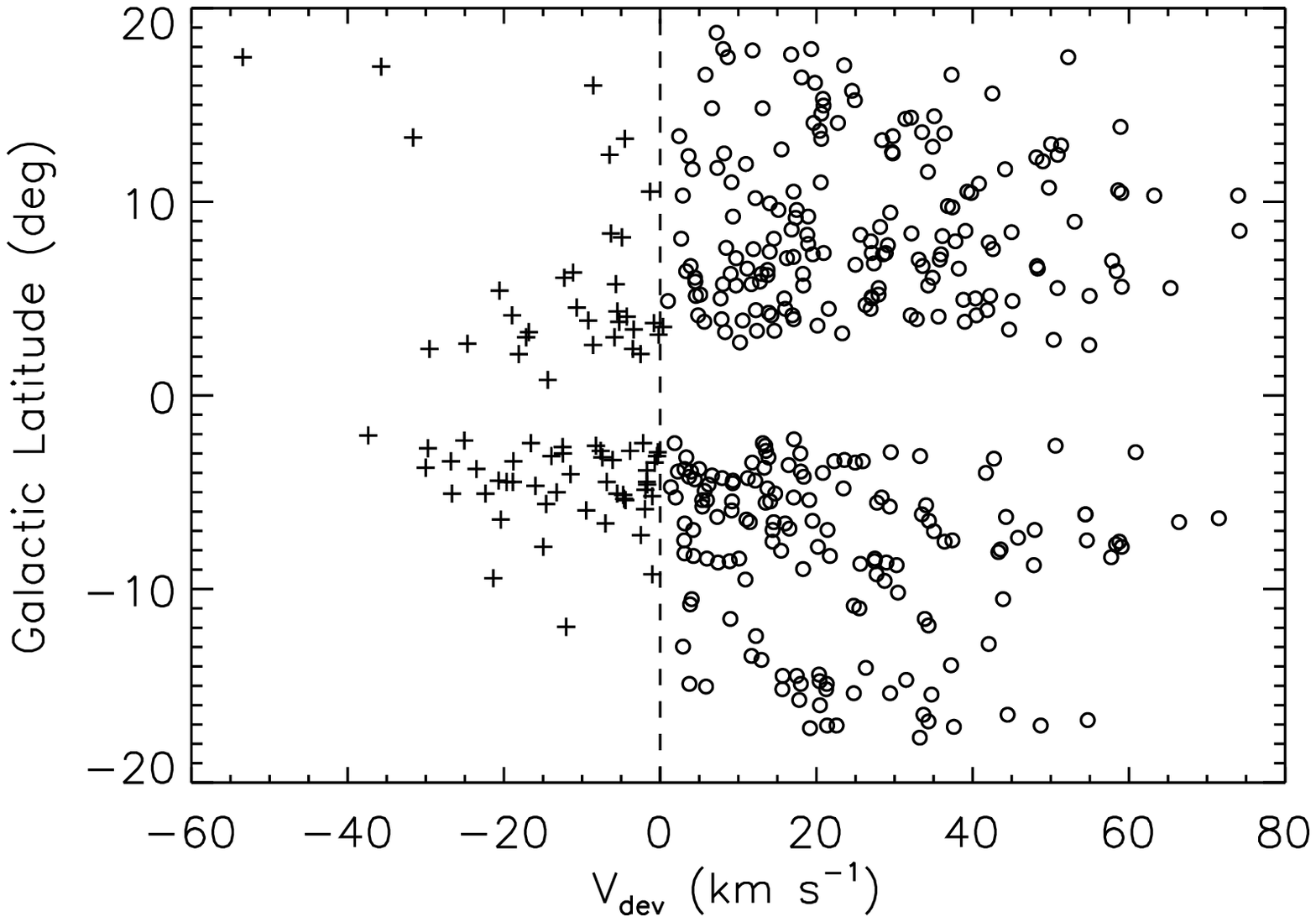}
\caption{\label{bvdev}Latitude distribution with deviation velocity,
  $V_{\mathrm{dev}}$, for all clouds in Table \ref{catalog}, where
  $V_{\mathrm{dev}}=V_{\mathrm{LSR}}-V_{\mathrm{t}}$, as defined by
  \citet{1991bWakker}. We used $V_{\mathrm{t}}$ determined by
  \citet{2007McClure-Griffiths} from \HI\ observations for $l \leq
  339.695\degr$ and by \citet{2006Luna} from CO observations for all
  remaining longitudes. Crosses denote a subset of clouds with 
  $V_{\mathrm{dev}}\leq 0.8$~km~s$^{-1}$ (the tangent
  point sample, \S\ref{tangentpoints}), while circles denote all other
  clouds in the GASS pilot region. For clouds that are observed at
  forbidden velocities ($V_{\mathrm{dev}}<0$~km~s$^{-1}$, denoted by
  the dashed line), those further from the plane have smaller absolute
  deviation velocities. This trend is expected of a population of
  clouds whose kinematics are dominated by Galactic rotation, as the
  number of clouds would fall off at more negative velocities and
  larger latitudes, but there are some outliers at large positive
  latitudes with more negative deviation velocities.} 
\end{figure}

In Figure \ref{bvdev} we display the latitude distribution with
deviation velocity. For clouds beyond the terminal velocity (with
negative deviation velocities), those at lower latitudes
have larger absolute deviation velocities. This is likely an artefact,
as the number of clouds would naturally
fall off with both more negative deviation velocities and larger
latitudes if they were dominated by Galactic rotation. There are, however, some outliers at large positive
latitudes with very negative deviation velocities.  

\begin{figure*}
\includegraphics[scale=0.36]{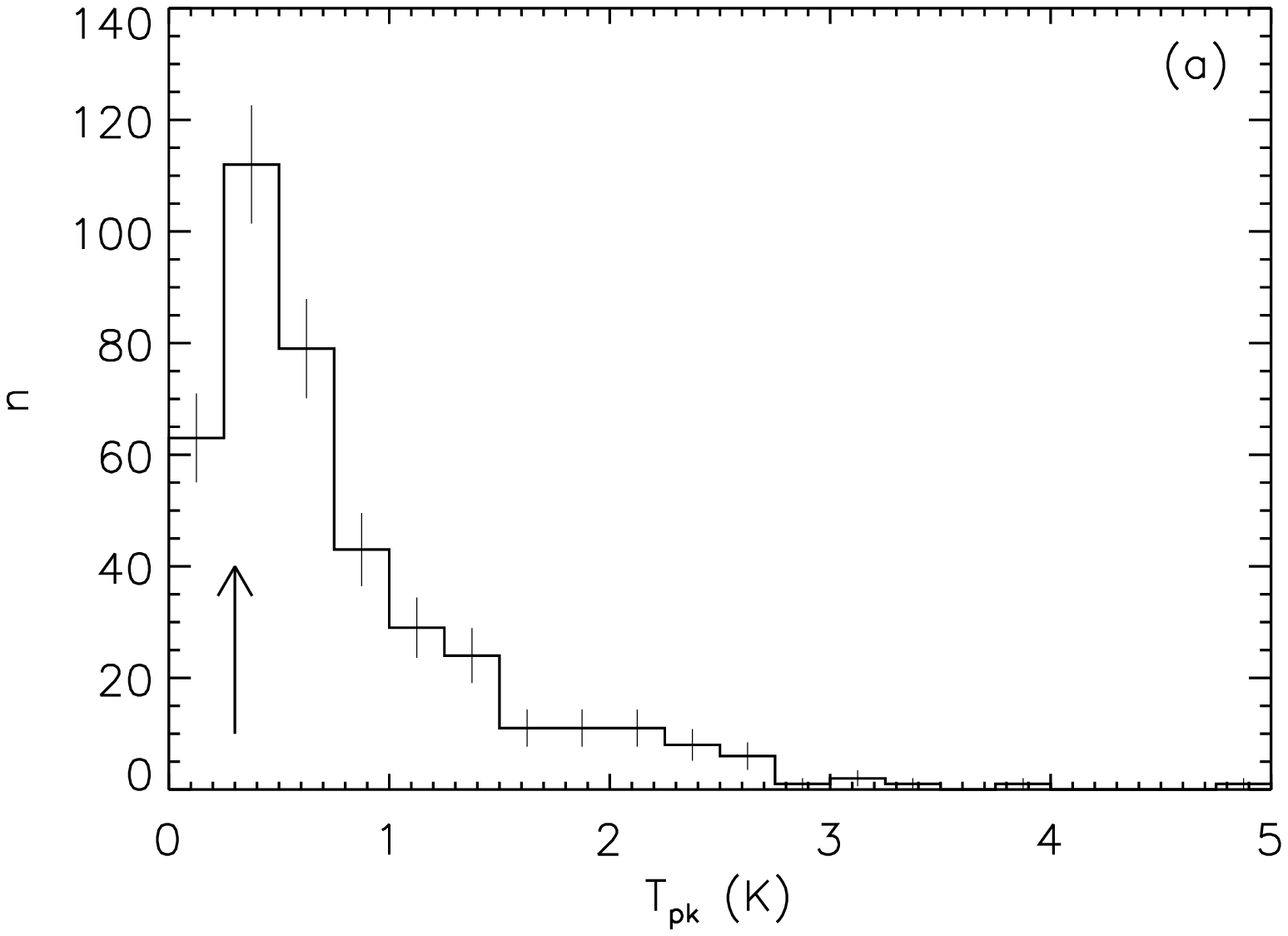}\qquad\includegraphics[scale=0.36]{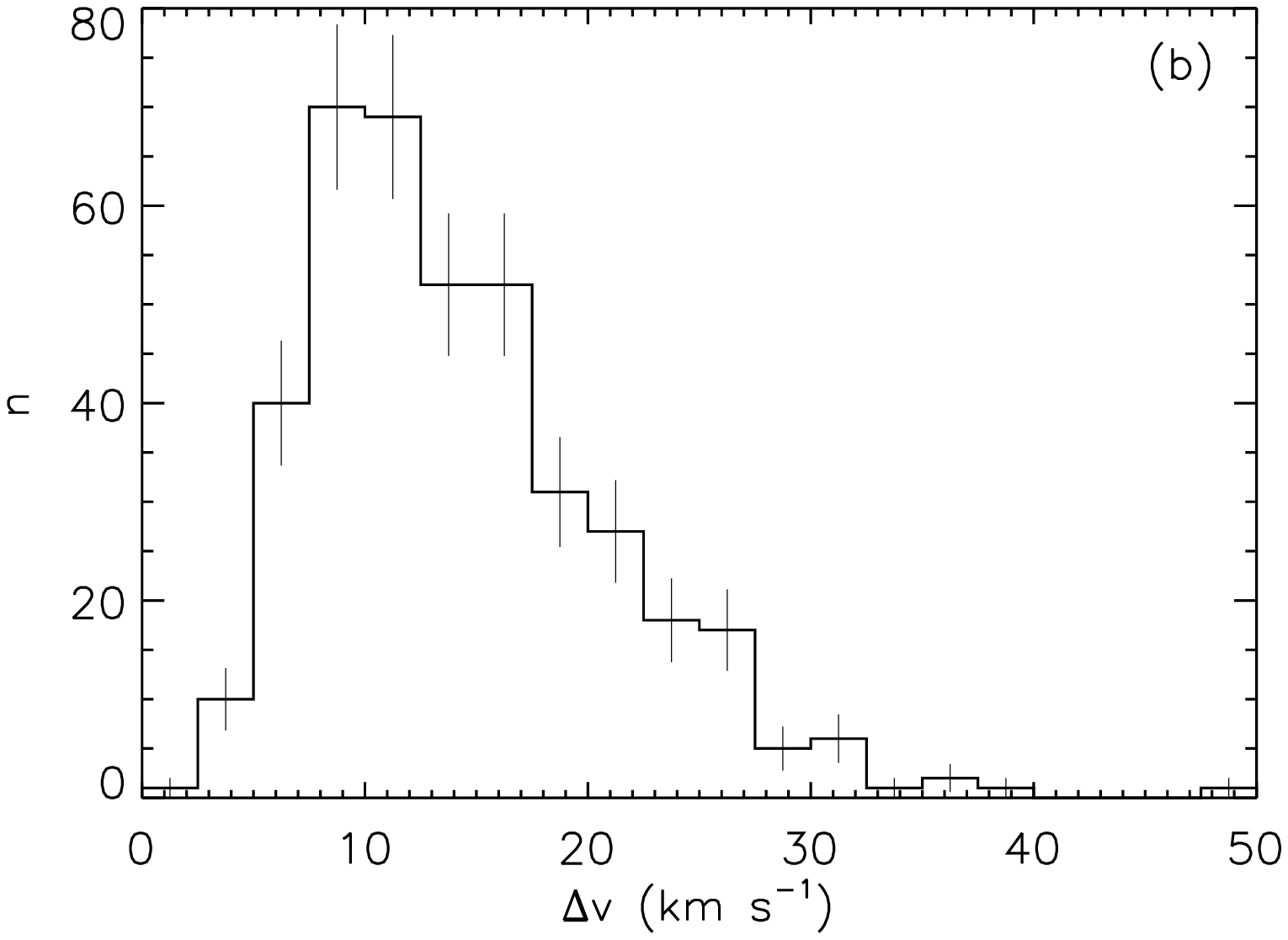}\qquad\includegraphics[scale=0.36]{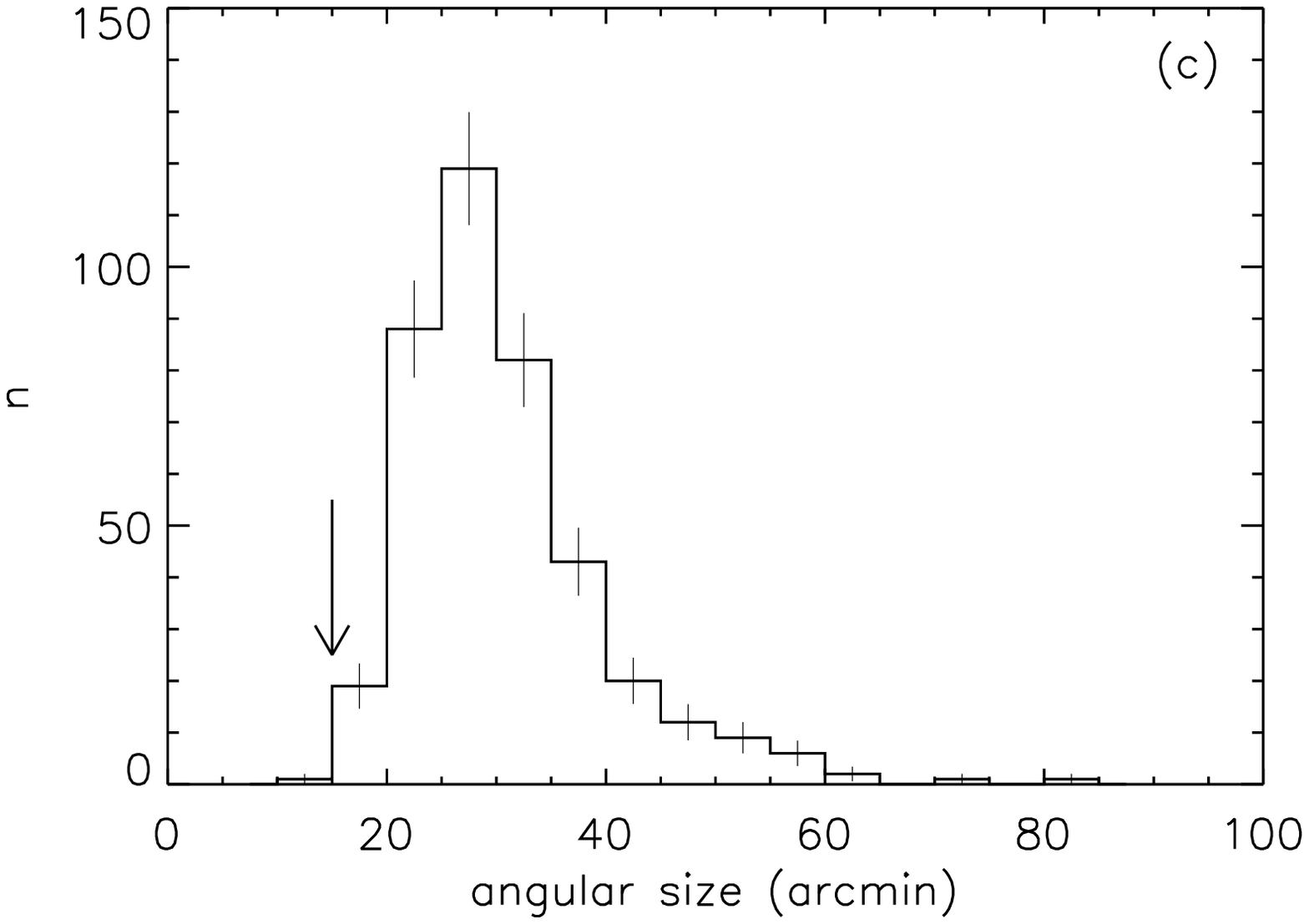}
\caption{\label{pTbhist}(a) Histogram of the peak brightness
  temperature of the clouds, $T_{\mathrm{pk}}$. The median is
  $T_{\mathrm{pk}}= 0.6$~K and the cutoff at small $T_{\mathrm{pk}}$
  is likely due to the sensitivity limit. The arrow represents the
  $5\Delta T_b$ detection level. We assume $\sqrt{N}$
  errors. (b)\label{fwhmhist} Histogram of the FWHM of the velocity
  profile of the clouds, where the median is $12.8$~km~s$^{-1}$. As
  the spectral resolution is $0.8$~km~s$^{-1}$ and all but one
  linewidth is greater than $3.4$~km~s$^{-1}$, the linewidths are well
  resolved. (c)\label{angularsizehist}  Histogram of the angular size
  of clouds, 
  $\sqrt{\theta_{\mathrm{maj}}\theta_{\mathrm{min}}}$, where
  $\theta_{\mathrm{maj}}$ and $\theta_{\mathrm{min}}$ are from Table
  \ref{catalog}. The median angular diameter of the clouds is $29'$,
  which is approximately twice the beam size. This median, along with
  the apparent lack of clouds with small angular sizes, is most likely
  due to the spatial resolution limit ($15'$, denoted by an arrow) and
  suggests that many of the clouds are unresolved.}
\end{figure*}

Histograms of $T_{\mathrm{pk}}$, $\Delta v$ and angular size are
presented in Figures \ref{pTbhist}a--c, respectively. The majority of
clouds have $T_{\mathrm{pk}}\leq 1$~K, with the median
$T_{\mathrm{pk}}=0.6$~K. The number of clouds decreases sharply below
$T_{\mathrm{pk}}\simeq 5\Delta T_b$ and suggests that the sample is
sensitivity limited. The median FWHM of the
velocity profiles is $12.8$~km~s$^{-1}$ and very few clouds have
linewidths larger than $30$~km~s$^{-1}$ (Figure \ref{fwhmhist}b). All
but one of the linewidths are greater than $3.4$~km~s$^{-1}$; as the
velocity resolution of the survey is $0.8$~km~s$^{-1}$, most lines are
therefore well resolved. The median angular diameter of the clouds is $29'$,
which is approximately twice the beam size. This value and
the steep cutoff at small angular sizes seen in Figure
\ref{angularsizehist}c are most likely due to the spatial resolution limit
of the data and suggest that many of the clouds are unresolved.

\section{Selection of a set of ``Tangent Point'' Clouds}
\label{tangentpoints}

The largest magnitude velocity from Galactic rotation in the inner
Galaxy occurs at the tangent point, 
defined as the location where the line-of-sight is
perpendicular to a circle of constant Galactocentric radius (see Figure
\ref{tangptdiagram}). Here $R_{\mathrm{t}}= R_0|\sin{l}|$ and the LSR
velocity from Galactic rotation is
$V_{\mathrm{t}}=R_0\left[\Theta/R_{\mathrm{t}}-\Theta_0/R_0\right]\sin{l}\cos{b}\mbox{,}$
where $R_0$  is the radius of the solar circle, $\Theta$ is the
circular velocity and $\Theta_0$ is the circular velocity at the solar
circle. We adopt $R_0\equiv 8.5$~kpc and $\Theta_0\equiv
220$~km~s$^{-1}$, as recommended by the IAU \citep{1986Kerr}. Clouds
in pure Galactic rotation cannot have a circular velocity beyond
$V_{\mathrm{t}}$. However, the random motion of a cloud near the
tangent point might increase the cloud's $|V_{\mathrm{LSR}}|$ beyond
$|V_{\mathrm{t}}|$. Clouds in
Galactic rotation with $|V_{\mathrm{LSR}}|\geq|V_{\mathrm{t}}|$ must
therefore lie near the tangent point and thus at a known distance,
$d_{\mathrm{t}}=R_0\cos{l}/\cos{b}$. While an assumption, it is
reasonable to adopt a distance of $d_{\mathrm{t}}$ for clouds with
$|V_{\mathrm{LSR}}|\geq|V_{\mathrm{t}}|$ given the rapid decline
in the number of clouds beyond the terminal velocity, as shown in
Figure \ref{lv}. Tangent point clouds constitute a sample uniquely
suited for investigating the population's distribution and properties,
such as physical size and mass. 

From the population of clouds detected in the GASS pilot region, we
define the tangent point sample as all clouds with
$V_{\mathrm{LSR}}\leq V_{\mathrm{t}}+0.8$~km~s$^{-1}$, where
$0.8$~km~s$^{-1}$ is one channel width, and assume that they are at
the tangent point; we assess the effect of this assumption in \S
\ref{disterrs}. We also assume that Galactic rotation is constant with
distance from the plane, i.e., $\Theta(z)=\Theta(0)$; deviations from
cylindrical rotation will be discussed in a subsequent work. As the
tangent point clouds in the GASS pilot region are nearly all at the
same distance and the latitude boundary of the region corresponds to a
constant height of $2.5$~kpc (at tangent points), they provide a
uniformly selected sample.    

\subsection{Derived Properties}
\label{derivedprops}

The physical properties and positions of the tangent point clouds are
presented in Table~\ref{catalogtangent}, while descriptions of the
derived quantities are presented below. 

Columns 1-3: As in Table \ref{catalog}.

Column 4: The deviation velocity,
$V_{\mathrm{dev}}=V_{\mathrm{LSR}}-V_{\mathrm{t}}$
\citep{1991bWakker}, where $V_{\mathrm{t}}$ is the most negative
velocity expected from Galactic rotation in the fourth quadrant, in
km~s$^{-1}$, and was determined by \citet{2007McClure-Griffiths}
from \HI\ observations for $l \leq 339.695\degr$ and by
\citet{2006Luna} from CO observations for all remaining longitudes.

Column 5:  The distance, $d$, along the line-of-sight from the
Sun to the cloud determined by assuming the cloud is at the tangent
point: $d=R_0\cos{l}/\cos{b}$, in kpc.  

Column 6: The Galactocentric radius, $R=R_0|\sin{l}|$, of the tangent
point at the cloud's location, in kpc.

Column 7: The height, $z$, of the cloud from the plane of the Galaxy,
determined geometrically to be $z=d\sin{b}$, in kpc. 

Column 8:  The radius, $r$, of the cloud in pc, determined by
$\sqrt{r_{\mathrm{maj}}r_{\mathrm{min}}}$, where
$r_{\mathrm{maj}}=(1/2)d\theta_{\mathrm{maj}}$,
$r_{\mathrm{min}}=(1/2)d\theta_{\mathrm{min}}$, and
$\theta_{\mathrm{maj}}$ and $\theta_{\mathrm{min}}$ are from
Table~\ref{catalog}.   

Column 9: The physical mass of \HI\ in the cloud, $M_{HI}$, determined
as in Column 8 of \S \ref{observedcolumns} but with the tangent point
distance assumed, in $M_\odot$.    

\subsection{Uncertainties in Derived Properties}
\label{deriveduncertainties}

Here we present error estimates for the derived properties in Table
\ref{catalogtangent}. 

$\delta V_{\mathrm{LSR}}$: As in Table \ref{catalog}.

$\delta V_{\mathrm{dev}}$: This error is $\sqrt{{\delta
    V_{\mathrm{LSR}}}^2+{\delta V_{\mathrm{t}}}^2}$, where $\delta
V_{\mathrm{t}}=3$~km~s$^{-1}$ for clouds located at longitudes where
    $V_{\mathrm{t}}$ was determined using \HI\ ($l \leq 339.695\degr$;
    \citealt{2007McClure-Griffiths}). For all other longitudes, where
    the terminal velocity was determined using CO observations
    \citep{2006Luna}, the error is assumed to be $9$~km~s$^{-1}$, as
    suggested from the scatter on Figure 7 of
    \citet{2007McClure-Griffiths}.   

$\delta d$: Distance errors, which are inherent to the assumption that
the clouds are located at tangent points, were estimated using a
simulated population of clouds (see \S \ref{disterrs}). 

$\delta R$: The error on the Galactocentric radius was determined
analogously to $\delta d$. Because the closest point to the Galactic
center along a given line-of-sight is the tangent point, the adopted
$R$ is  always a lower limit and the error can only be positive.

$\delta z$: The error on the height is estimated to be $\delta d|\sin{b}|$.

$\delta r$: The error on the radius of the cloud depends on the
uncertainties in the distance and angular size estimates and is
$\delta r=r\sqrt{{(\delta \theta/\theta)}^2+{(\delta d/d)}^2}$, where
$\theta=\sqrt{\theta_{\mathrm{maj}}\theta_{\mathrm{min}}}$ \\
and $\delta \theta=(1/2)\sqrt{{\delta \theta_{\mathrm{min}}}^2\left(\theta_{\mathrm{maj}}/\theta_{\mathrm{min}}\right)+{\delta \theta_{\mathrm{maj}}}^2\left(\theta_{\mathrm{min}}/\theta_{\mathrm{maj}}\right)}$.

$\delta M_{HI}$: The error on the mass is
    $M_{HI}\sqrt{(\delta M_{HI}d^{-2}/M_{HI}d^{-2})^2+(2\delta
    d/d)^2)}$, where $M_{HI}d^{-2}$ is given in Table \ref{catalog}.  

\section{ANALYSIS OF THE TANGENT POINT CLOUD POPULATION}
\label{discussion}

\subsection{Simulated Halo Cloud Population}
\label{simulations}

\begin{figure}
\plotone{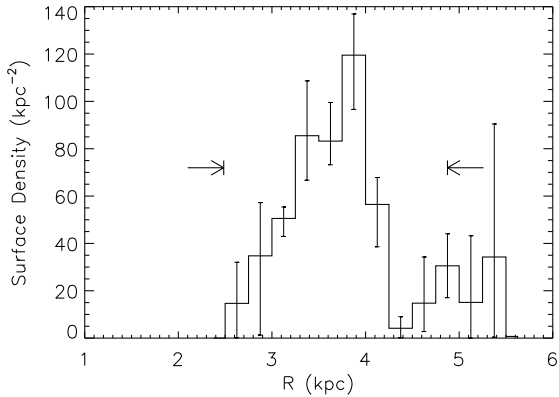}
\caption{ \label{Rhist} Best estimate of the radial surface density,
  $\Sigma(R)$, of clouds in the GASS pilot region. Error bars denote the
range in amplitude within each bin for the three radial solutions.
 Clouds were detected at all longitudes within the GASS pilot region, where
  corresponding tangent point boundaries in $R$ are indicated by arrows. The
  distribution suggests that the clouds are concentrated in radius,
  peaking at $R=3.8$~kpc.} 
\end{figure}

To constrain the spatial and kinematic properties of the observed
tangent point cloud population we simulated a population of
clouds to which were applied the same $l$, $b$ and $V_{\mathrm{LSR}}$
selection criteria as for the GASS pilot region clouds.
The simulated clouds were randomly sampled from the following
distribution: 
\begin{equation}
n(R,z) = \Sigma(R)\exp\left[{-\frac{|z|}{h}}\right],
\end{equation}
where $\Sigma(R)$ is the radial surface density distribution, $h$ is
the exponential scale height and $R$ and $z$ are the cylindrical
coordinates. $\Sigma(R)$ is composed of $12$ independent radial bins
of width $0.25$~kpc, spanning $R=2.5$ to $5.5$~kpc. The amplitude
of each bin was optimized to best fit the observed longitude
distribution of the tangent point clouds by minimizing the Kolmogorov-Smirnov (K-S) $D$ statistic
(the maximum deviation between the cumulative distributions) using
Powell's algorithm \citep{1992Press}. We optimized the fits using
three different initial estimates on $\Sigma(R)$. All converged on a
similar solution and we adopted the mean of the three as the best fit
to the data, which is shown in Figure \ref{Rhist}.

Velocities of simulated clouds were based on a
flat rotation curve where $\Theta=\Theta_0$ with a random velocity
component drawn from a Gaussian of dispersion $\sigma_{cc}=18$~km~s$^{-1}$, which is
discussed in detail in \S\ref{veldispersion}. We generated 
$5\times 10^4$ clouds in a half-galaxy (third and fourth quadrants), 
of which $2475$ were within
the Galactic coordinates of the GASS pilot region and also within the
defined velocity range of the tangent point cloud sample. We then normalized
this distribution to compare directly with the observed
distribution. We performed K-S tests to estimate
the quality of the fit between the observed and simulated
distributions. Based on these tests, we find that the parameters of
the simulated population reproduce the distribution of observed
tangent point clouds well and are therefore good estimates for those
of the intrinsic population. Results of the fits to the
distributions are presented in \S \ref{veldispersion}--\ref{verthist}
along with tests of other functional forms. 

\subsection{Cloud-to-Cloud Velocity Dispersion}
\label{veldispersion}

\begin{figure}
\plotone{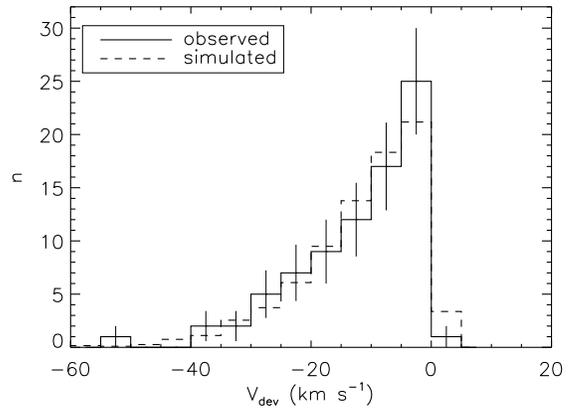}
\caption{ \label{simvdev} Distribution of deviation velocities of
  the observed tangent point and simulated
  population of clouds, where the simulated clouds are normalized to
  the same total number of observed tangent point clouds. Observed
  clouds are represented by a solid line, while simulated clouds are
  represented by a dashed line.  To
  reduce the effects of small number statistics, simulations were run
  with $5\times 10^4$ clouds. We assume $\sqrt{N}$ errors.}
\end{figure}

We assume that a random cloud-to-cloud velocity dispersion
($\sigma_{cc}$) is responsible for the presence of clouds at forbidden
velocities within the GASS pilot region. Simulations of the global
cloud population were required to model the effects of these motions,
which can cause clouds that are not located at tangent points
to have velocities that are near or beyond $V_{\mathrm{t}}$. The
cloud-to-cloud velocity dispersion can provide information on the
expected scale height of the distribution and a better understanding
of the formation mechanisms of the clouds. We find the
$V_{\mathrm{dev}}$ distribution of the tangent point sample of clouds
to be consistent with that derived from a Gaussian distribution of
random velocities whose dispersion is $\sigma_{cc}=18$~km~s$^{-1}$, with a K-S test probability greater than $97\%$%
\footnote{For reference, a $97\%$ probability is roughly equivalent in
  confidence to a $0.04\sigma$ detection of a difference, i.e.,
  no detectable difference in the distributions, $15\%$ corresponds to
  $1.4\sigma$, and $1\%$ corresponds to $2.6\sigma$.}. 
These distributions are presented in Figure
\ref{simvdev}, where the simulated $V_{\mathrm{dev}}$ distribution is
represented by a dashed line and the observed distribution by a solid
line. Velocity dispersions of $16$ to $22$~km~s$^{-1}$ also provide
acceptable fits (K-S test probabilities greater than or equal to
  $15\%$), as do fits where the random velocity component is drawn
  from an exponential distribution rather than a Gaussian distribution
  (with a K-S test probability of $89\%$ for a scale velocity of
  $13$~km~s$^{-1}$).
The implied kinematics of the cloud population based on this result are
discussed in \S \ref{kinematics}.

Estimating the uncertainties on our determined $\sigma_{cc}$ is
extremely difficult. We expect these uncertainties to be coupled to
errors introduced by the measurements of $V_{\mathrm{t}}$, which are
difficult to untangle due to the possibilty that random motions can
systematically offset the measured $V_{\mathrm{t}}$. Another possible
source of uncertainty is streaming motions associated with spiral
features. However, because $V_{\mathrm{t}}$ was determined directly
from \HI\ and CO measurements rather than from a fitted rotation
curve, streaming motions should already be reflected in the adopted
$V_{\mathrm{t}}$. 

\subsection{Distances Errors}
\label{disterrs}

\begin{figure}
\plotone{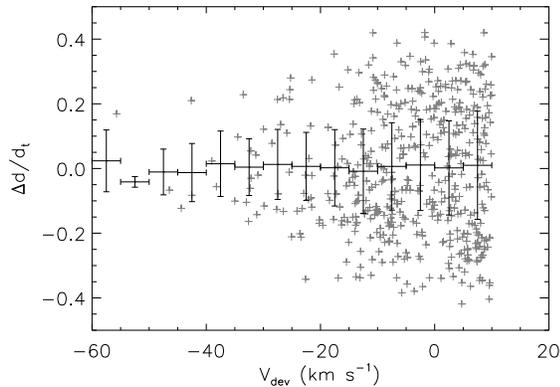}
\caption{\label{disterrors} Relative distance error as a function of
  deviation velocity (crosses) for a  population of simulated
  clouds. Clouds with a large forbidden velocity (i.e., large negative value
  of $V_{\mathrm{dev}}$) are more likely to be at the tangent point
  and thus have smaller errors in their estimated distances.
  These points have been placed into bins of width 
  $5$~km~s$^{-1}$, denoted by the horizontal bars, and the
  vertical error bars denote the mean and standard
  deviation of the distribution within each bin. Approximately one
  fifth of simulated points are displayed here but the error bars have
  taken all simulated data into account. The left most error bar is
  large due to small number statistics.}
\end{figure}

The simulations allow us to estimate the error in our assumption that
clouds with $V_{\mathrm{dev}}\leq 0.8$~km~s$^{-1}$ are at the tangent
point. We have calculated the fractional distance error of the
simulated clouds as a function of deviation velocity within a
$5$~km~s$^{-1}$ wide $V_{\mathrm{dev}}$ bin (Figure
\ref{disterrors}). Clouds at increasingly forbidden velocities have
smaller distance errors, i.e., are more likely to be near the tangent
point; the degree to which this is true depends on the magnitude of
$\sigma_{cc}$. We also calculated the fractional distance error as a
function of longitude and found that clouds with longitudes
corresponding to the Galactocentric radii where few clouds are
detected ($l \sim 328\degr$) have larger distance errors (this is
expected because few clouds are intrinsically at these radii and
therefore a larger fraction of the forbidden velocity clouds at these
longitudes are likely interlopers from larger radii). We assume that
the errors due to $V_{\mathrm{dev}}$ and longitude are
independent. The adopted fractional distance error is the product of
the fractional distance error due to $V_{\mathrm{dev}}$ (relative to
the typical error of $0.12$), due to longitude (relative to the
typical error) and the typical error itself. The fractional  distance
errors due to $V_{\mathrm{dev}}$ and longitude are taken to be the rms
error for the simulated clouds within the same bin. We have confirmed
that the fractional distance error of the simulated clouds does not
significantly depend on their latitude. Based on the relative distance
errors, we believe that our assumption that clouds with
$V_{\mathrm{dev}} \leq 0.8$~km~s$^{-1}$ are located near their tangent
point is reasonable.      
  
\subsection{Radial Distribution}
\label{radhist}

\begin{figure}
\plotone{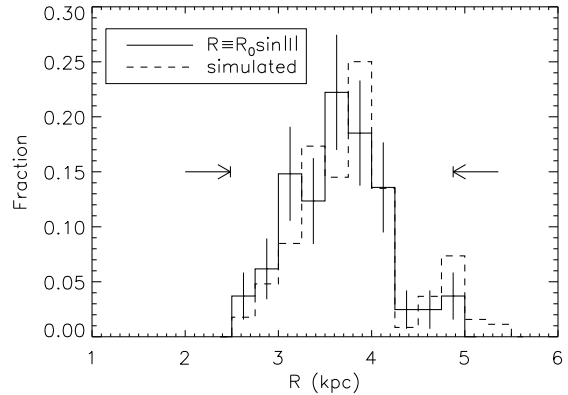}
\caption{\label{Rerrors} Histogram showing the fraction of the sample
  located at each Galactocentric radius. The solid line denotes the
  clouds assumed to be located at tangent points and the dashed line
  denotes the corresponding
  simulated radial distribution (dashed line). The shift between the
  simulated and adopted radial distributions results from systematic
  errors that stem from the assumption that the forbidden velocity
  clouds are located at tangent points, i.e., at the minimum possible
  value of $R$. Arrows represent the radial boundaries of tangent
  points within the GASS pilot region. We assume $\sqrt{N}$ errors.}  
\end{figure}

The adopted radial distribution of the tangent point clouds along with
that of the simulated population of clouds is shown in Figure
\ref{Rerrors}. The apparent offset in the distributions results from
the assumption that the forbidden velocity clouds are located at
tangent points, which are always at the smallest value of $R$ along the
line of sight.

\begin{figure*}
\plottwo{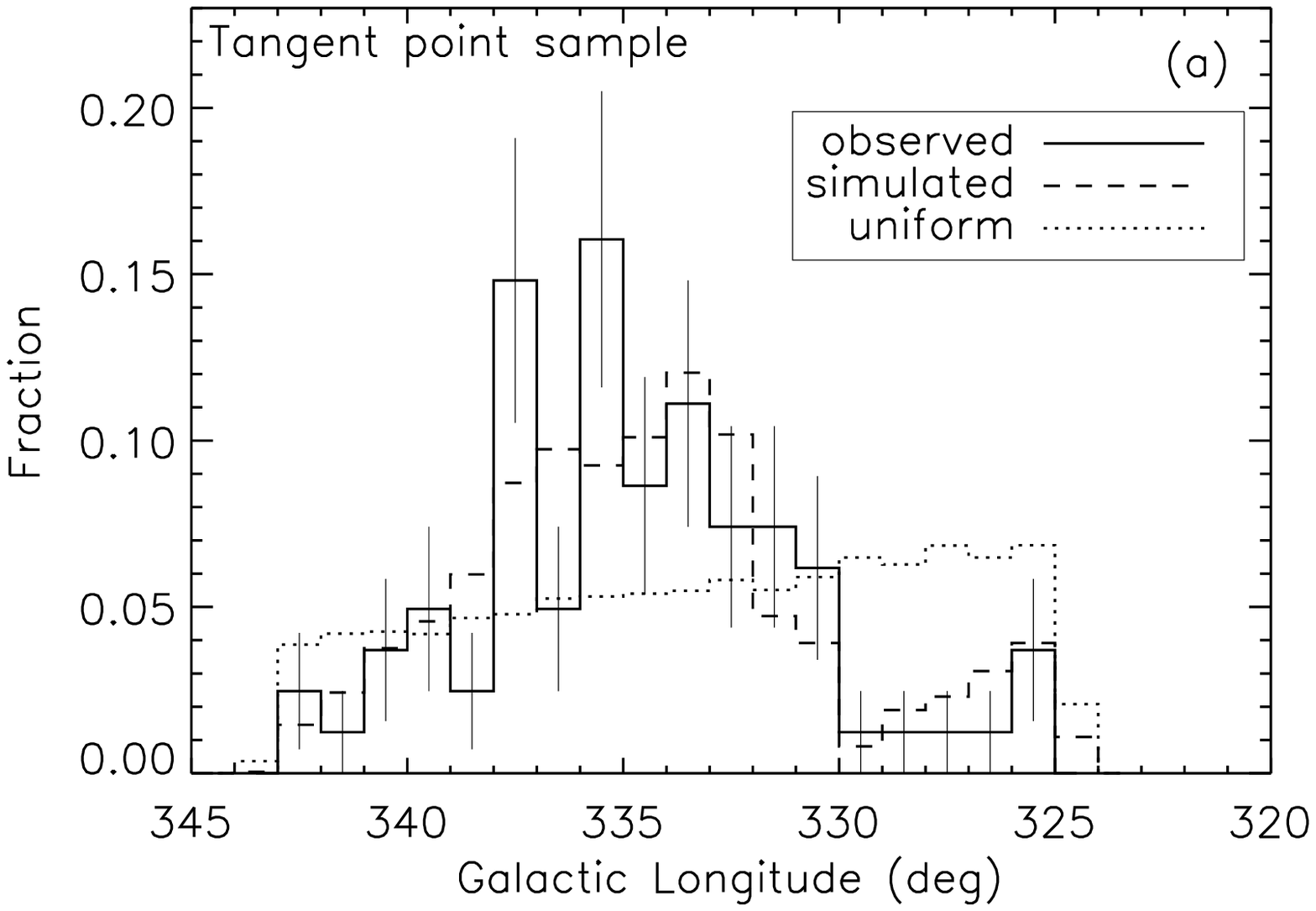}{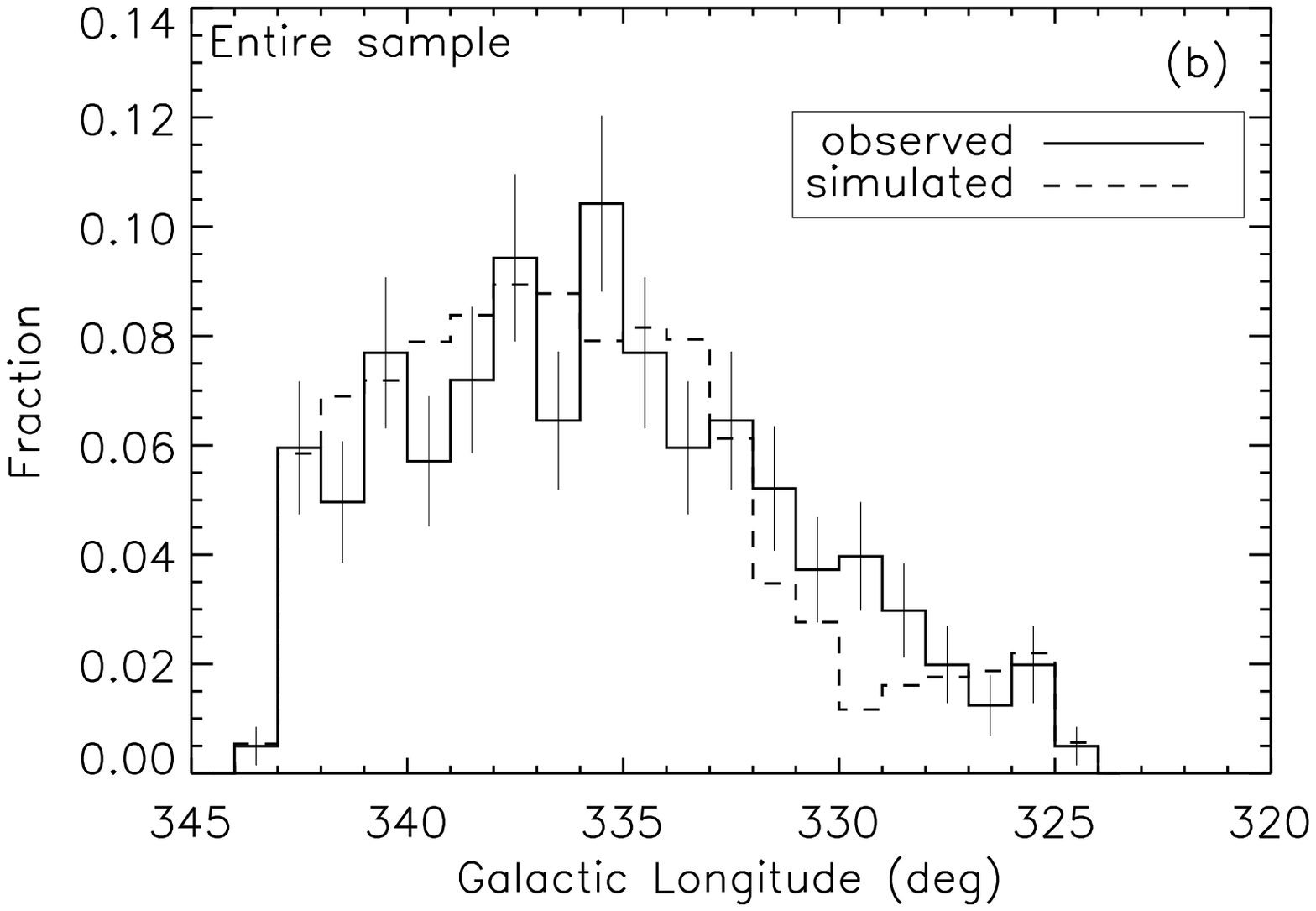}
\caption{
  \label{lhist}(a) Histogram presenting the longitude distribution of
  the observed tangent point clouds (solid line) and corresponding
  simulated distribution (dashed line), along with the distribution 
  expected from a uniform
  population (dotted line). There is a peak in the observed and
  simulated distribution, which is clearly different from that
  expected of a uniform population. (b) Histogram presenting the longitudes of all
  clouds detected in the GASS pilot region (solid line) along with
  the corresponding simulated population (dashed line). The
  distributions are similar, suggesting that the tangent point sample
  represents the pilot region fairly well. We assume $\sqrt{N}$ errors.} 
\end{figure*}

Another useful quantity that can be extracted from the simulations is
the mean radial surface density distribution of clouds within the GASS
pilot region (Figure \ref{Rhist}). Although clouds are observed at all
longitudes within the GASS pilot region, the distribution is
concentrated in Galactocentric radius and peaks at $R=3.8$~kpc. The
error bars at $R\gtrsim 5$~kpc are significantly larger because clouds
at these radii must have large random velocities in order to meet the
sample criteria and therefore represent a small tail of the
population. With a K-S test probability of $95\%$, the simulated
longitude distribution fits that of the observed distribution of the
tangent point sample well (Figure \ref{lhist}a). 

The number of clouds
in a uniformly distributed population of tangent point clouds should
monotonically decrease by a factor of $\sim 2$ between $l=325\degr$
and $l=343\degr$, as demonstrated by Equation (A2) in
\citet{2006Stil}, where they show that the line-of-sight distance
effectively surveyed over forbidden velocities is $\Delta
d=\sqrt{8}R_0|\sin{l}|\sqrt{\sigma_{cc}/\Theta_0}\propto|\sin{l}|$. We
have overlayed the distribution of a uniform surface density
population in Figure \ref{lhist}a and it is clear that it is in
stark contrast to the centrally peaked longitude distribution we observe. We
therefore conclude that the peaked radial distribution is real. In Figure
\ref{lhist}b we present the longitude distributions of the simulated
and observed clouds within the entire pilot region, i.e., at all
$-200 \le V_{\mathrm{LSR}} \le -70$ km~s$^{-1}$.
Even though the vast majority of these clouds were not used to
constrain the simulated radial distribution, their longitude
distribution is well reproduced. The tangent point sample therefore
appears to be a fair representation of the entire GASS pilot region.

It is worth investigating whether our assumptions regarding the
functional form of the velocity distribution affects the inferred
radial distribution, i.e., could a distribution with more velocity
outliers and a less strongly peaked radial distribution also fit the
data? To test this we have performed the same optimization of the
radial distribution while using an exponential velocity
distribution. We find that the resulting radial profile is identical
to within the errors, confirming the robustness of this result.

\subsection{Vertical Distribution}
\label{verthist}

\begin{figure}
\plotone{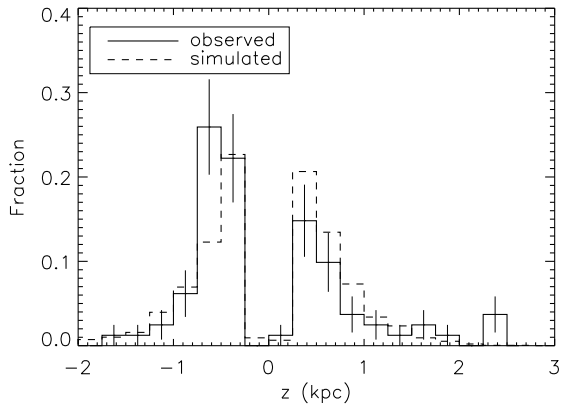}
\caption{ \label{zhist} Histogram of the fraction of the observed (solid
  line) and simulated (dashed line) clouds as a function of
  height. The lack of observed clouds at low $|z|$ is a selection
  effect due to confusion. This effect was taken into account during
  comparisons with the simulated population by omitting all simulated
  clouds with $|b|\leq 2\degr$. The exponential scale height of the
  distribution is $h=400$~pc. We assume $\sqrt{N}$ errors.} 
\end{figure}

The vertical distribution of both the tangent point and simulated
cloud population is presented in Figure \ref{zhist}. Clouds have been detected
throughout the entire range of latitudes covered by the GASS pilot
region, up to corresponding heights of $|z|=2.5$ kpc. However, there
are very few clouds at $|b|\leq 2.5\degr$ because identification of
clouds was extremely difficult close to and within the Galactic plane,
except in cases where clouds were observed at large forbidden
velocities.
Because of this incompleteness at low latitudes, we compare
the latitude distributions of simulated and observed clouds with
$|b|>2\degr$ and find that they are consistent for vertical scale
heights between $400-500$~pc, with K-S test probabilities of at least
$25\%$. As incompleteness may still be a problem at $|b| \approx
2\degr$, we also compare the height distributions for only clouds at
$|b|>3\degr$ and find acceptable fits for scale heights between $300$
and $400$~pc, both with probabilities greater than $50\%$. Based on
these comparisons we conclude that the population is best represented
with a exponential scale height of $h=400$~pc. A $\mathrm{sech}^2(z/z_0)$
distribution with $z_0=700$~pc provides an equally good fit the the
data, which is not surprising given that the exponential and $\mathrm{sech}^2$
distributions differ primarily near $z=0$~pc where we cannot
constrain the fit.

Although incomplete, the combination of surface density, mean mass and
scale height of the clouds gives a rough estimate of the vertical
distribution of \HI\ contributed by the cloud population. The
mid-plane \HI\ number density can be estimated by
$n(0)=\Sigma(R)\left<M\right>/(2hM_{\mathrm{H}})$, where
$\left<M\right>$ is the mean cloud \HI\ mass and $M_{\mathrm{H}}$
is the \HI\ atom mass. We find that the clouds are responsible for
$\sim 5\%$, by \HI\ number density, of the exponential component of
the \HI\ layer in \citet{1990Dickey} and have a very similar scale height.

We note that there is an asymmetry in the number of observed clouds
above and below the disk at $|z|\leq750$~pc and an excess of clouds
at large positive latitudes. Possible explanations for these are discussed
in \S \ref{nature}.

\subsection{Physical Size and Mass}
\label{mass}

\begin{figure*}
\plottwo{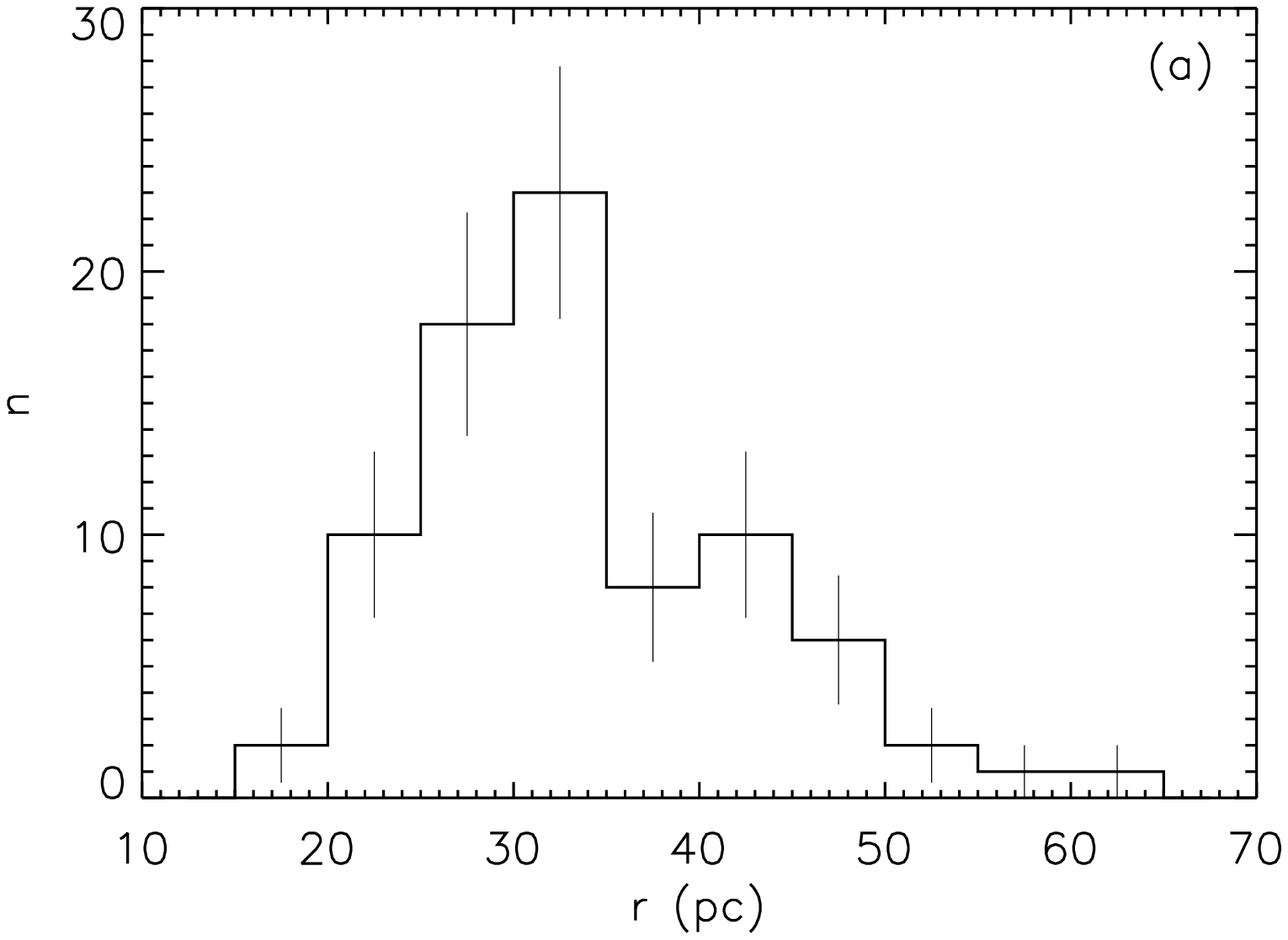}{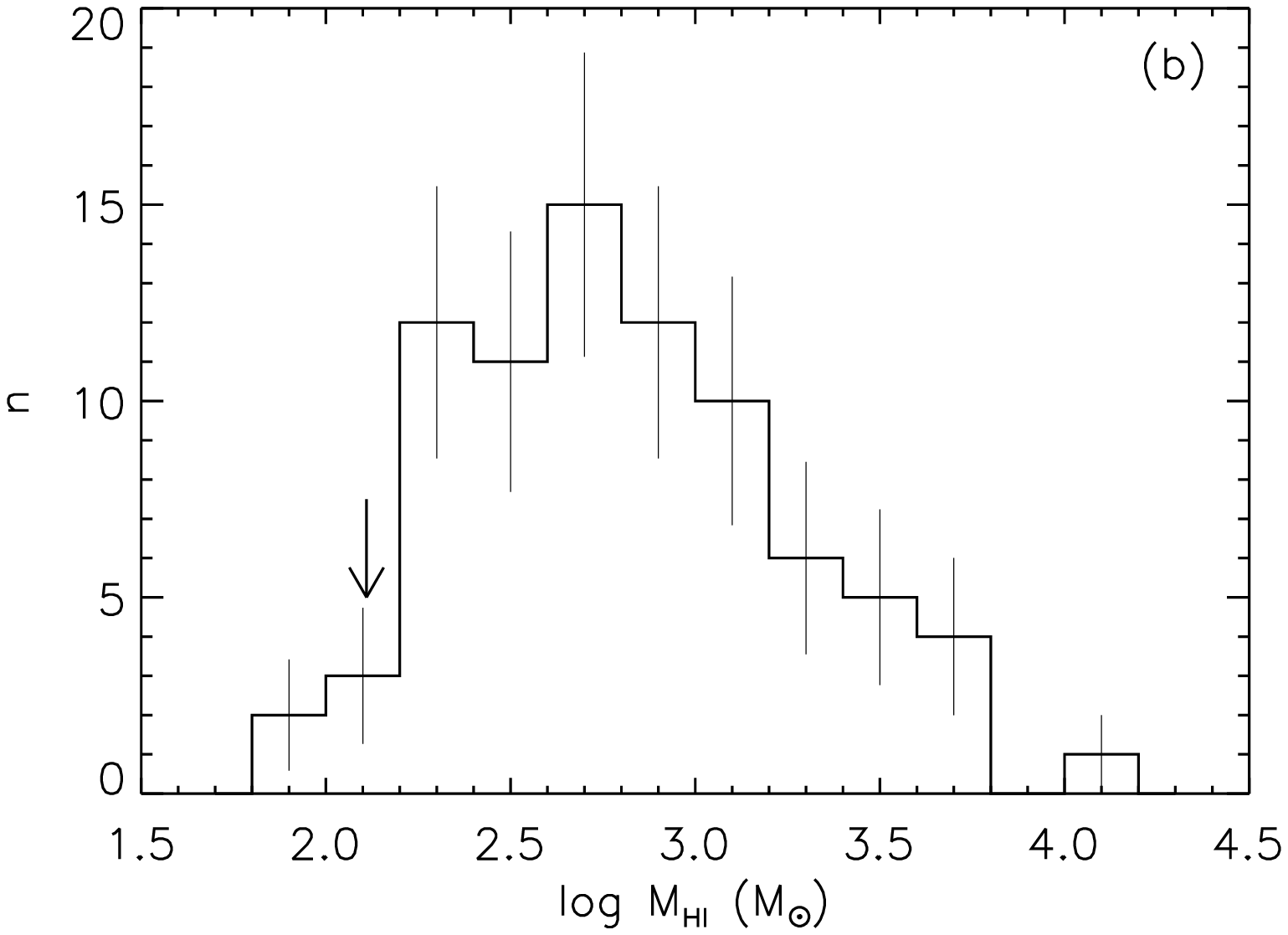}
\caption{ (a) \label{radiushist} Histogram of the
  physical radius of clouds at tangent points. The median
  $r=32$~pc. The cutoff at small radii is a result of the spatial
  resolution limit. (b) \label{masshist} Histogram of log $M_{HI}$ of
  clouds at tangent points. The $\sim 5\Delta T_b$ limit is
  represented by an arrow for a cloud with the median $r=32$~pc. We
  assume $\sqrt{N}$ errors.}
\end{figure*}

Cloud radii, $r$, vary from $\sim 15$ pc to $65$ pc, with a median
radius of $32$ pc, as can be seen in Figure \ref{radiushist}a. The
angular resolution of the telescope sets a lower limit on the observed
cloud size. The maximum angular extent of the detected clouds suggests
that roughly $80\%$ of the entire sample is unresolved in at least
one dimension. Derived radii should therefore be thought of as
upper limits. A histogram of the \HI\ mass of the clouds at tangent
points is presented in Figure \ref{masshist}b, demonstrating that the
clouds range in size from hundreds to thousands of solar masses, with
a median \HI\ mass of $630~M_\odot$. These values may be overestimates
if confusion is significant.

\subsection{Comparison of GASS Clouds to Lockman Clouds}
\label{comparisons}

Based on their angular sizes and location in the lower halo, the
clouds detected in the GASS pilot region appear to be similar to those
observed by \citet{2002Lockman}. We investigate this possibility further by
comparing the properties of each distribution summarized in
Table~\ref{comptable}. The median $\Delta v$ is strikingly similar in
both sets of data, which is not surprising if the clouds
are part of the same
population. The derived $\sigma_{cc}$ are also in agreement. The
median $|z|$ of this sample is smaller than that of
\citet{2002Lockman} but this is most likely due to a selection effect,
as the areas searched by \citet{2002Lockman} tended to be further from
the Galactic plane to avoid areas of confusion. The most obvious
differences in the sample stem from the difference in the angular
resolutions of each survey: $15\arcmin$ for GASS data versus
$9\arcmin$ for the GBT data. This affects $T_{\mathrm{pk}}$, $N_{HI}$, and $r$, where $T_{\mathrm{pk}}$ and $N_{HI}$ would naturally
be lower for unresolved clouds, and $r$ would be
larger. If confusion is important, $M_{HI}$ would also be larger. \citet{2002Lockman} estimated that $25\%$ of the clouds were
unresolved while we have estimated $\sim 80\%$ here. The clouds
observed by \citet{2002Lockman} are much less massive, with
approximately one third having $M_{HI} \leq 30 M_\odot$, while none of
the observed GASS clouds have masses that low. As the only differences
in the observed properties are due to differences in the observations,
these comparisons reveal that the clouds belong to the same population
of clouds as those detected by \citet{2002Lockman}. 

\subsection{Observed Trends}

\begin{figure}
\plotone{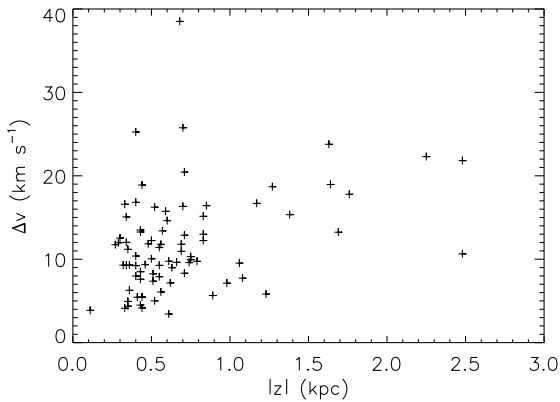}
\caption{\label{zfwhm} FWHM of the velocity profile, $\Delta v$, of the tangent
  point clouds as a function of height. Clouds at larger heights
  may have larger linewidths.}  
\end{figure}

There does not appear to be any correlation between the height of the
clouds and $V_{\mathrm{LSR}}$, $R$, or $r$. However, as the data in
Figure \ref{zfwhm} suggest, there may be a trend between $\Delta v$
and $|z|$, where clouds near the plane ($|z|\leq 1$~kpc) have a median
FWHM of $10$~km~s$^{-1}$ and a large dispersion, where as those at
$|z|> 1$~kpc have a median FWHM of $17$~km~s$^{-1}$ and a smaller
dispersion. One possible explanation for such a trend could be that
the clouds at larger heights belong to a different population of
clouds than those at lower heights.
Another possibility is that if the clouds are in pressure
equilibrium, the trend is reflecting pressure variations throughout the
halo. Similar to our results, \citet{2002Lockman} found that
$V_{\mathrm{LSR}}$ and $r$ are independent of $z$, and also found
evidence that clouds with more narrow linewidths lay closer to the plane.

\begin{figure}
\plotone{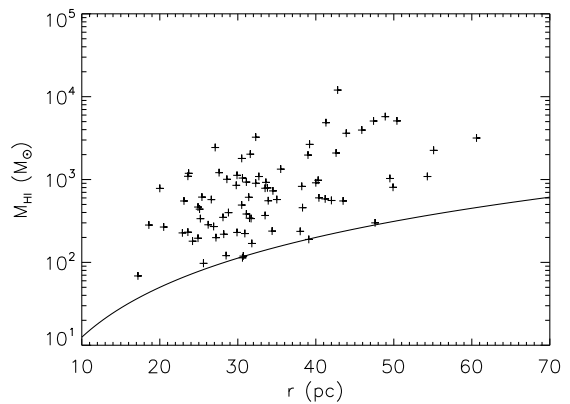}
\caption{\label{mradius} Derived \HI\ mass of the clouds as a function
  of cloud radius (crosses). The curved line
  represents the lower \HI\ mass limit and is based on the minimum
  observable $N_{HI}$. The minimum brightness temperature is assumed
  to be $5\Delta T_{\mathrm{b}}$.  The larger clouds are more massive.}   
\end{figure}

As one might expect if a population of clouds has a narrow range of
densities, the larger the radius of the cloud, the more massive it
is. This trend is evident in Figure \ref{mradius} and does not appear
to be solely due to the \HI\ mass sensitivity limit of the
data. This limit is denoted by the curved line and is based on the
minimum observable \HI\ column density, $N_{HI_\mathrm{min}}=1.94\times
10^{18}T_{b_\mathrm{min}}\Delta v_{\mathrm{med}}$ in cm$^{-2}$, where
$T_{b_\mathrm{min}}$ is the minimum observable $T_b$ (assumed to be
$5\Delta T_b$) and $\Delta v_{\mathrm{med}}$ is the
observed median $\Delta v$. 

\begin{figure}
\plotone{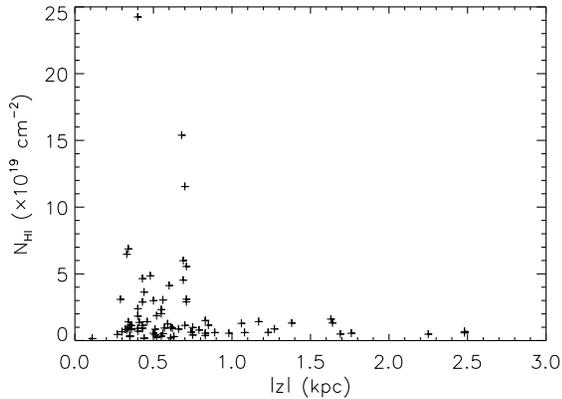}
\caption{\label{zNHI} \HI\ column density as a function of
  height. All clouds with large column densities are located at lower
  heights, i.e., closer to the Galactic disk.} 
\end{figure}

\begin{figure}
\plotone{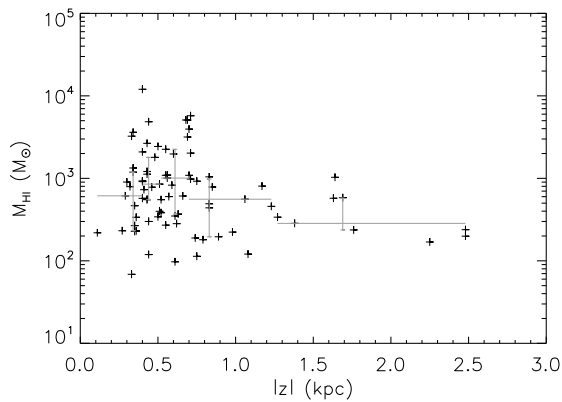}
\caption{\label{mz} \HI\ mass of the clouds as a function of
  height (crosses). Horizontal lines represent the median $M_{HI}$
  per $18$ clouds and error bars represent the $25^{\mathrm{th}}$ to
  $75^{\mathrm{th}}$ percentile range. There is no evidence for a
  trend until $|z|\sim 1$~kpc, where there may be a lack of massive
  clouds.} 
\end{figure}

\begin{figure}
\plotone{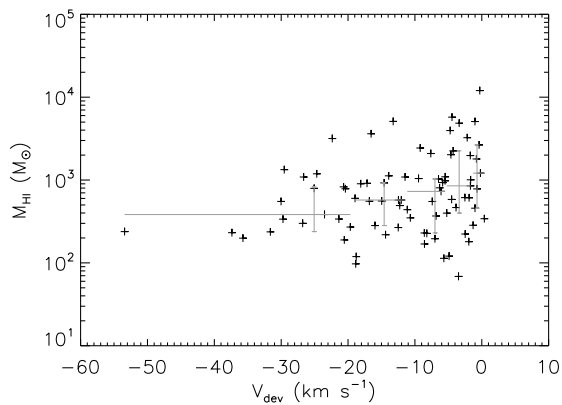}
\caption{\label{mvdev} \HI\ mass of the clouds as a function of
  deviation velocity (crosses). Horizontal lines represent the median $M_{HI}$
  per $18$ clouds and error bars represent the $25^{\mathrm{th}}$ to
  $75^{\mathrm{th}}$ percentile range. The more massive clouds have
  lower deviation velocities.} 
\end{figure}

Another apparent trend, as seen in Figure \ref{zNHI}, is that the clouds
with higher \HI\ column densities appear to be at lower
heights. This could be due to inclusion of unrelated diffuse emission
with the clouds at lower heights if the
background subtraction was not effective. It could also be explained
by a scenario where each cloud was given the same kinetic energy from
a formation process or via equipartition in the subsequent evolution
of the cloud population. The higher mass clouds, which have higher
$N_{HI}$, would then have preferentially lower velocities and would
not reach heights as large as those reached by clouds with higher
velocities. Figure \ref{mz} provides tentative support for this
hypothesis, revealing that at $|z|\geq 1$~kpc there are very few
clouds with $M_{HI}\geq 10^3M_{\odot}$ (note, however, that at these
heights our statistics are poor). \citet{2002Lockman} did not find a correlation between
the \HI\ column density or mass with height, but this could be due to
the limited vertical range of his data. Similarly, a correlation between the \HI\ mass of a cloud and its
deviation velocity is suggested in Figure \ref{mvdev}; in particular,
the more massive clouds may have lower deviation velocities. If the
more massive clouds have lower random motions, they could have lower
typical deviation velocities. This would be consistent with a scenario
in which clouds evolve to an equipartition of energy or in which each
cloud is given a similar initial kick, for example, from similar
supernovae explosions. This could also explain why all of the most
massive clouds are seen closer to the plane (Figure \ref{mz}): if they
have smaller initial velocities, they would not move as far into the
halo before falling back towards the plane. At this stage, however, we
can not exclude the possibility that at lower heights confusion is
affecting the determined \HI\ column density and mass of the clouds. 
We discuss the kinematics of the cloud population further in \S \ref{kinematics}.  

\section{The Origin and Nature of Halo Clouds}
\label{nature}

\subsection{Kinematics of Halo Cloud Population}
\label{kinematics}

The velocity dispersion of the cloud population is likely a remnant of
a common formation process, such as violent supernovae explosions. If
the population is in equilibrium, the vertical distribution and
velocity dispersion are linked via the potential. By using the
simulated population of clouds, we found that the vertical
distribution is best fit with an exponential scale height of $400$~pc
(\S\ref{verthist}) and the $V_{\mathrm{dev}}$ distribution is best fit
with a Gaussian dispersion of $\sigma_{cc}=18$~km~s$^{-1}$
(\S\ref{veldispersion}). However, assuming a vertical force using the
mass model of \citet{2007Kalberla} at $R=3.8$~kpc, we derive an 
exponential scale height of $90$~pc for an isothermal population of
clouds with $\sigma_{cc}=18$~km~s$^{-1}$. To produce our observed
scale height of $400$~pc within this potential, the cloud-to-cloud
velocity dispersion would have to be $60$~km~s$^{-1}$. The difference
between the $\sigma_{cc}$ required and that observed may be due to the
lack of clouds observable in the disk; if a large number of clouds
within the disk have gone undetected, the scale height could be lower
than $400$ pc, and could therefore be explained by the observed
velocity dispersion. It is also possible that the distribution cannot
be explained by a single component, but this is not obvious in the
current data (we will address this possibility in a subsequent
paper). However, the magnitude of this difference suggests that the
clouds do not belong to an equilibrium population and their heights
must, in part, result from processes that do not increase the velocity
dispersion, such as uniformly expanding \HI\ shells, instead of bursts
of energy from areas of active star formation generating random
``kicks.'' The clouds could have also originated above the disk, as in
a galactic fountain model.  

Similar clouds have been detected at forbidden velocities within the
Galactic disk using data from the VLA Galactic Plane Survey,
suggesting that the clouds are not restricted to the halo
\citep{2006Stil}. Those cloud diameters are much smaller than the
ones derived here (they have diameters $\sim 10$~pc),
likely due to the higher spatial resolution of the VLA data. Based on
their models, \citet{2006Stil} find that a vertical Gaussian half-width at
half-maximum (HWHM) larger than $1$ kpc (equivalent to an exponential
scale height of $1.2$~kpc) best fits their data, but given that they
only surveyed within $|b|\leq 1.3\degr$, this is likely to be strongly
affected by the lack of coverage at high latitudes, especially because
we observe similar clouds up to $|z|=2.5$ kpc and \citet{2002Lockman}
has observed clouds up to $\sim1.5$ kpc. \citet{2006Stil} find a lower limit
to the HWHM of the clouds to be $180$~pc (exponential scale height of
$216$~pc), which is consistent with the value we derive and also
inconsistent with the value expected for the derived velocity
dispersion. 

It is worth noting that there is evidence that similar clouds may
also be abundant in the outer Galaxy \citep{2006Stanimirovic}. It is
not yet certain whether they belong to the same population of clouds,
but with larger surveys of \HI\ clouds in the lower halo of the
Galaxy, such as those presented here and in the entire inner Galaxy
within GASS, along with those underway by the Galactic Arecibo L-Band
Focal Plane Array Consortium and other groups, properties
such as $\sigma_{cc}$ and the spatial distribution can be determined
more accurately and will therefore help constrain the kinematics and
formation mechanisms of the clouds.   

\subsection{Halo Clouds and Spiral Structure}
\label{spiralarms}

As discussed in \S \ref{radhist}, the surface density of halo clouds
is not uniformly distributed but instead peaks at a Galactocentric
radius of $R=3.8$ kpc and the population is concentrated in radius. This confined nature
suggests that the halo clouds are related to the spiral structure of the
Galaxy. Although distances to spiral arms are currently not well
constrained, the location of the peaked radial distribution of the
clouds indicates that they may be related to the expanding ``$3$~kpc''
arm \citep{1957vanWoerden,1960Rougoor}, which is tangential to an
observer's line-of-sight at roughly $l=336\degr$
(\citealt{2008Bronfman}; see also \citealt{2008Vallee} who argues that
this feature lies at $l\sim 339\degr$ and is the start of the Perseus
arm, distinct from the $3$~kpc--Norma arm), corresponding to
$R=3.5$~kpc. Also, it has been suggested that the $3$~kpc arm must be
confined to an annulus of less than $1$ kpc in extent
\citep{1981Lockman}, which corresponds well with the radial
concentration of clouds. At this time, however, the
possibility that these clouds are instead related to other Galactic
structures such as the ``$5$~kpc molecular ring'' \citep{2004Jackson}
cannot be ruled out. This ``ring'' is likely not a coherent structure
but instead a complex region where multiple spiral arms originate
\citep{2008Vallee}.     

\begin{figure}
\plotone{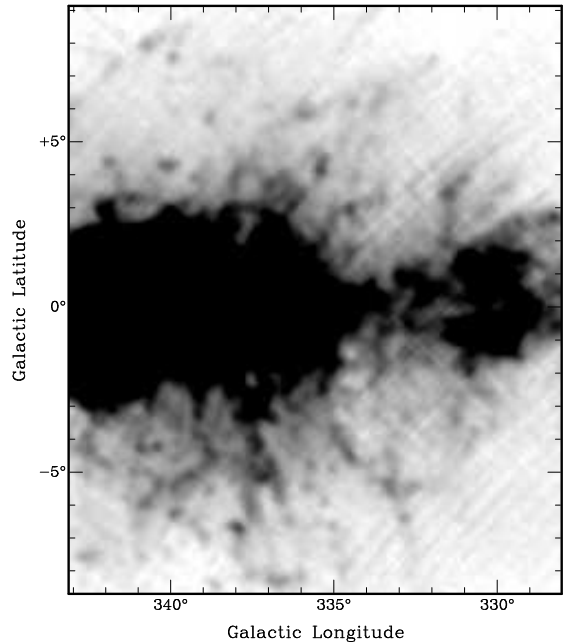}
\caption{\label{newshell} Filamentary structures are plentiful within
  the GASS pilot region, as demonstrated in this longitude--latitude
  integrated intensity map over $ -135 \leq V_{\mathrm{LSR}}\leq
  -120$~km~s$^{-1}$. Many \HI\ clouds in the GASS pilot region are
  aligned with these loops and filaments, suggesting that they are
  related to expanding superbubbles or other structures common in areas
  of star formation. The diagonal striations are instrumental artifacts.} 
\end{figure}

The apparent association between many of the \HI\ clouds and
filamentary structures within the GASS pilot region suggests that the
clouds are related to star formation because such structures are
common in areas of significant supernova activity or stellar winds
\citep{1990Dickey}. Figure \ref{newshell} displays an integrated
intensity map with many clouds that are clearly aligned along 
filaments and loops, reminiscent of the clouds observed to be
associated with the cap of a superbubble by
\citet{2006McClure-Griffiths}. Recently, proper motion measurements of
the molecular cloud NGC 281 West, which is associated with an \HI\
loop, revealed that the cloud is moving away from the Galactic plane at
$20-30$~km~s$^{-1}$ \citep{2007Sato}. This velocity is
similar to our observed velocity dispersion and further supports the
scenario that the clouds are related to expanding shells. In this scenario, violent
supernovae and stellar winds may have pushed \HI\ from the disk up
into the halo or the clouds may be fragments of \HI\ shells
\citep{1989MacLow,2006McClure-Griffiths}, rather than a result of
a standard galactic fountain. In the galactic fountain model hot gas
rises from the disk, cools and condenses, then falls back to the
plane \citep{1976Shapiro,1980Bregman}. This would result in a more
uniform radial distribution of clouds \citep{1980Bregman},
while we clearly observe a peak in the radial distribution of the
clouds that may be associated with the $3$~kpc arm. If the halo
clouds are related to star formation, the asymmetry in the number of
detected clouds at low heights ($|z|\leq 750$~pc) could be a result of
this, as any asymmetry in the structure of the ISM and the location of
star forming regions may be reflected in the distribution of clouds,
and there appears to be more filaments below than above the Galactic
plane in the GASS pilot region.   

\begin{figure}
\plotone{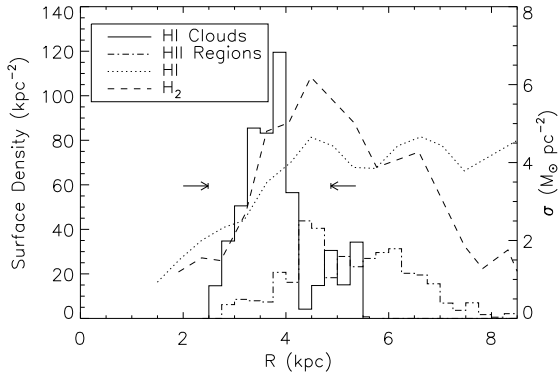}
\caption{\label{H2} Radial surface density distribution of the
  simulated GASS pilot region \HI\ clouds (solid histogram) and \HII\
  regions (scaled up by a factor of $7$ for ease of comparison and
  shown by the
  dash-dot histogram; \citealt{2004Paladini}). Mass surface densities of the
  average Galactic \HI\ (dotted line) and Galactic  $\mathrm{H}_2$
  (dashed line) are overlaid and were both determined by
  \citet{1993Dame} using data from \citet{1990Dickey} and
  \citet{1978Burton} for \HI\, and data from \citet{1988Bronfman} for
  $\mathrm{H}_2$. Arrows indicate limits of $R$ for tangent points in the GASS pilot region.  
  The peaked distribution of halo clouds does not appear to be similar
  to that of the \HII\ regions nor the \HI\ and $\mathrm{H}_2$
  mass surface densities, suggesting that the halo clouds are not
  directly related to current star formation.}
\end{figure}

If the clouds are related to spiral structure and star formation then
we would expect to see a correlation between the radial surface
density distribution
of the \HI\ clouds and that of Galactic \HII\ regions. We compare
these distributions, along with the mass surface densities of \HI\ and
$\mathrm{H}_2$, in Figure \ref{H2}. The mass surface densities have been
averaged over the entire Galaxy and were derived by \citet{1993Dame}
using \HI\ data from \citet{1990Dickey} and \citet{1978Burton}, and
$\mathrm{H}_2$ data from \citet{1988Bronfman}, whereas the halo cloud
distribution from this study includes only 
the GASS pilot region. The \HII\ regions are taken
from \citet{2004Paladini}. There is no obvious relationship between
the \HI\ cloud and \HII\ region distributions or between the \HI\
clouds and the \HI\ and $\mathrm{H}_2$ surface densities, providing
conflicting evidence for the relationship between the \HI\ clouds and
current star formation. As the evaporation timescale for a cold cloud
in a hot medium is expected to be much longer than other timescales
associated with cloud evolution \citep{1977Cowie,2006Nagashima},
and if evaporation is the main disruptive mechanism, the clouds
are long-lived. It is therefore possible that the clouds are tracing
past rather than current star formation. With the present data we are
only able to make a preliminary study of the relationship between the
halo cloud distribution and spiral structure and star formation in the
Galaxy. Future analysis of clouds over a larger range of longitudes
will allow us to test the assertion that the clouds are related to
star formation in greater detail.  

\subsection{Possible Association with High Velocity Cloud Complex L}
\label{complexLsection}

An excess of clouds at large positive heights can be seen in the
vertical distribution of the clouds (Figure \ref{zhist}); $9\%$ of the
clouds at positive heights lie at $z\geq 2$~kpc while none of the
clouds below the plane are seen at such heights, and many of them have
unusually large deviation velocities (Figure \ref{bvdev}). This excess
could have several possible origins including infalling gas, increased
disk activity on one side of the disk that has resulted in outflowing
gas reaching larger heights, or small number statistics. Given the
proximity of high velocity cloud complex L to the halo clouds in the
upper portion of the GASS pilot region, we compared the \HI\
associated with each population to determine whether or not the
presence of complex L could be responsible for the observed excess. 

Complex L was first described by \citet{1991Wakker} to have velocities
ranging from $-190 \lesssim V_{\mathrm{LSR}} \lesssim -85$ km
s$^{-1}$, longitudes ranging from $341\degr \leq l \leq 348\degr$ and
latitudes ranging from $31\degr\leq b \leq 41\degr$, and they
speculated that the clouds were part of
a population that was related to a galactic fountain. Since then,
H$\alpha$ distance limits have been determined for complex L, placing
it within the Galactic halo at heights of $4\leq z\leq 12$~kpc from
the plane and heliocentric distances of $8$ to $22$~kpc
\citep{2001Weiner,2003PutmanB}.   

\begin{figure}
\plotone{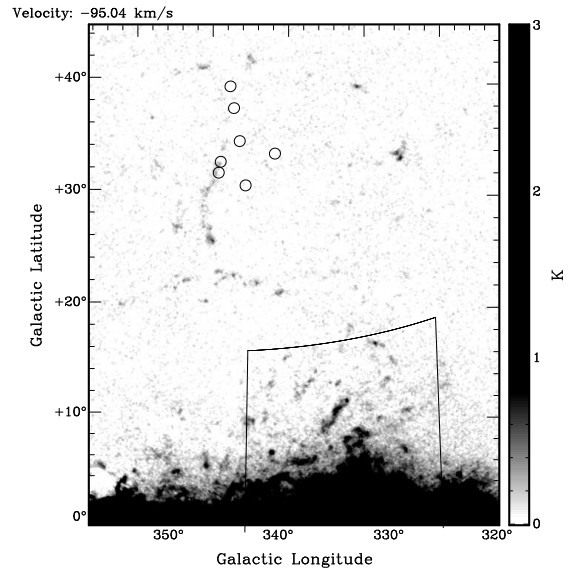}
\caption{\label{complexL} GASS data encompassing the upper portion of
  the GASS pilot region, outlined by the solid black lines, along with
  the high velocity cloud complex L. Circles have been placed at the
  positions of high velocity clouds within complex L, regardless of
  their $V_{\mathrm{LSR}}$, as determined by \citet{1991Wakker}. The
  morphology of the \HI\ in the upper GASS pilot region, in complex L, and
  between them, suggests that complex L may have similar origins as 
  the observed halo clouds and may be responsible for the observed
  excess of clouds at large positive latitudes. }  
\end{figure}

In Figure \ref{complexL} we display a region of the GASS data that
encompasses both the upper GASS pilot region, outlined by the solid
black lines, and the lower velocity gas of complex L. We have overlaid
circles on high velocity clouds associated with complex L as
catalogued by \citet{1991Wakker}, regardless of their velocity.
The spatial morphology of the \HI\ at velocities where gas is detected
in complex L suggests that there is a connection between the clouds in
the GASS pilot region and those in complex L, in the form of a
filamentary structure that is contiguous in velocity and connects
complex L with the disk. Also, with the assumption that
the halo clouds at large positive latitudes are located at tangent
points, their heights are between $1.5\leq z \leq 2.5$ kpc with
distances $\sim 8$~kpc, which place them in the vicinity of the lower
height estimates of complex L. These correlations suggest that complex
L may have similar origins as the clouds presented here and it may
be responsible for the observed excess of clouds at large positive
latitudes.  

\subsection{Stability of Halo Clouds}
\label{stability}

The amount of mass required for a spherical cloud to be
gravitationally bound is $M\approx r\Delta v^2/G$, where $r$ is the
radius of the cloud, $\Delta v$ is the linewidth of its velocity
profile and $G$ is the gravitational constant. For a cloud with
$r=32$~pc and $\Delta v=12.8$~km~s$^{-1}$, the median observed values,
this would require a mass on the order of $10^6 M_{\odot}$. As none of
the detected clouds have masses this large, they are either pressure
confined or transitory.  

Because turbulence also contributes to the linewidths, an upper limit
for the thermal pressure within the clouds is $P=nT \leq
22\left<n\right>\Delta v^2$, in units of K cm$^{-3}$,  where
$\left<n\right>$ is the average number density and $T$ is the thermal
temperature. However, as $\left<n\right>$ is a lower limit and $T$ is
an upper limit, and $\left<n\right>$ and $T$ are uncorrelated, we are
unable to put any constraints on the pressure of these clouds with the
present data. It is possible that pressure changes are responsible for
the observed trend in Figure \ref{zfwhm}, which tentatively shows that clouds
further from the disk of the Galaxy have larger linewidths. If the
clouds are in pressure equilibrium, their pressures could provide us
with insight into the pressure structure of the halo, which is
currently not well understood.      

\section{SUMMARY}
\label{conclusions}

We have detected over $400$ \HI\ clouds in the lower halo of the
Galaxy in the Galactic All-Sky Survey pilot region. These clouds have
a median peak brightness temperature of $0.6$~K, a median velocity
width of $12.8$~km~s$^{-1}$, and angular sizes
$\lesssim 1\degr$. As these clouds follow Galactic rotation, a
subset was selected that is likely to be located at tangent points of the inner
Galaxy, allowing us to determine their distances and therefore their
sizes and masses. The tangent point clouds have radii on the order of
$30$~pc and a median \HI\ mass of $630 M_{\odot}$. The properties of
these clouds suggest that they belong to the same population of clouds
discovered by \citet{2002Lockman}.  

We simulated the population of clouds to constrain their random
cloud-to-cloud velocity dispersion, $\sigma_{cc}$, and spatial
distribution. We found that $\sigma_{cc}=18$~km~s$^{-1}$, but if the
clouds were left to evolve in the Galactic potential without any 
disruptions, these random motions would produce
a scale height of $90$~pc, which is inconsistent with our derived
scale height of $400$~pc. This suggests that the clouds do not belong to
an equilibrium population. We detected clouds throughout the entire
GASS pilot region, up to the latitude boundaries ($|b|\lesssim
20\degr$). Few clouds were observed at low latitudes due to confusion,
which may have resulted in an underestimate of the number of clouds at
low heights, and therefore an overestimate in the derived scale
height.          

Our large, homogeneously-selected sample has allowed us to determine
the spatial distribution of these halo clouds for the first time and
has revealed that although clouds were observed at all longitudes
within the GASS pilot region, they do not appear to be uniformly
distributed but instead are concentrated in radius, peaking at
$R=3.8$~kpc. We analyzed this distribution
and suggest that the clouds are related to the spiral structure of the Galaxy.
In particular, the peak in the radial distribution is suggestive of a
relation to the $3$~kpc arm. This relation to a specific spiral
feature remains speculative until further analysis using a larger
sample of clouds and better constrained spiral structure models can be
performed. It is therefore unlikely that the halo clouds are a
result of a standard galactic fountain, as radial enhancements
would not be expected in this scenario \citep{1980Bregman}. Instead, it
appears that the clouds may be directly related to areas of active
star formation, in the form of fragmenting \HI\ shells and \HI\ gas
that has been pushed into the halo. This is further supported by the
appearance of numerous clouds related to filaments and loops, whose
structures may have resulted from stellar winds and supernovae
\citep{1990Dickey}. However, a comparison between the radial surface
density distribution
of \HI\ clouds and \HII\ regions provides conflicting evidence: if
clouds are related to areas of star formation, a relationship between
the clouds and \HII\ regions would be expected but is not observed.    

The morphology of \HI\ at large positive latitudes in the GASS pilot
region suggests that some of the clouds may be related to high
velocity cloud complex L, whose lower height estimates of $4$~kpc and
distance estimates of $8$~kpc place it in the vicinity of the clouds
presented here. If they are related, it is possible that whatever
process is responsible for the halo clouds, e.g., star formation, is
also responsible for complex L. This is further supported by recent 
observations of
IVCs, which suggest that the IVCs are
related to energetic events in the Galactic disk and that they are
likely linked to spiral structure \citep{2006Kerton}. The majority of
those IVCs have $V_{\mathrm{dev}}\leq 20$~km~s$^{-1}$, implying  a
Galactic origin, and we suggest that they are similar to the cloud
population presented here. 

A future study of halo clouds within the entire inner Galaxy observed
by the Galactic All-Sky Survey will provide a much larger sample of
tangent point clouds, enabling more complete statistics for
distribution studies and a better understanding of Galactic
structure. These studies will include comparisons between the
distribution of halo clouds and that of tracers of star formation,
such as \ion{H}{2} and $\mathrm{H}_2$, and will allow us to better
constrain the origin of these clouds.

\acknowledgements
The Parkes Observatory is operated by the Australia Telescope National
Facility, a division of the Commonwealth Scientific and Industrial
Research Organisation. 

{\it Facilities:} \facility{Parkes ()}

\bibliography{ms.bib}

\begin{thebibliography}{54}
\expandafter\ifx\csname natexlab\endcsname\relax\def\natexlab#1{#1}\fi

\bibitem[{{Barnes} {et~al.}(2001){Barnes}, {Staveley-Smith}, {de Blok},
  {Oosterloo}, {Stewart}, {Wright}, {Banks}, {Bhathal}, {Boyce}, {Calabretta},
  {Disney}, {Drinkwater}, {Ekers}, {Freeman}, {Gibson}, {Green}, {Haynes}, {te
  Lintel Hekkert}, {Henning}, {Jerjen}, {Juraszek}, {Kesteven}, {Kilborn},
  {Knezek}, {Koribalski}, {Kraan-Korteweg}, {Malin}, {Marquarding}, {Minchin},
  {Mould}, {Price}, {Putman}, {Ryder}, {Sadler}, {Schr{\"o}der}, {Stootman},
  {Webster}, {Wilson}, \& {Ye}}]{2001Barnes}
{Barnes}, D.~G. {et~al.} 2001, \mnras, 322, 486

\bibitem[{{Bregman}(1980)}]{1980Bregman}
{Bregman}, J.~N. 1980, \apj, 236, 577

\bibitem[{{Bronfman}(2008)}]{2008Bronfman}
{Bronfman}, L. 2008, \apss, 313, 81

\bibitem[{{Bronfman} {et~al.}(1988){Bronfman}, {Cohen}, {Alvarez}, {May}, \&
  {Thaddeus}}]{1988Bronfman}
{Bronfman}, L., {Cohen}, R.~S., {Alvarez}, H., {May}, J., \& {Thaddeus}, P.
  1988, \apj, 324, 248

\bibitem[{{Burton} \& {Gordon}(1978)}]{1978Burton}
{Burton}, W.~B., \& {Gordon}, M.~A. 1978, \aap, 63, 7

\bibitem[{{Cowie} \& {McKee}(1977)}]{1977Cowie}
{Cowie}, L.~L., \& {McKee}, C.~F. 1977, \apj, 211, 135

\bibitem[{{Dame}(1993)}]{1993Dame}
{Dame}, T.~M. 1993, in American Institute of Physics Conference Series, Vol.
  278, Back to the Galaxy, ed. S.~S. {Holt} \& F.~{Verter}, 267

\bibitem[{{de Avillez}(2000)}]{2000deAvillez}
{de Avillez}, M.~A. 2000, \mnras, 315, 479

\bibitem[{{Dickey} \& {Lockman}(1990)}]{1990Dickey}
{Dickey}, J.~M., \& {Lockman}, F.~J. 1990, \araa, 28, 215

\bibitem[{{Ferri{\`e}re}(2001)}]{2001Ferriere}
{Ferri{\`e}re}, K.~M. 2001, Reviews of Modern Physics, 73, 1031

\bibitem[{{Heiles}(1967)}]{1967Heiles}
{Heiles}, C. 1967, \apjs, 15, 97

\bibitem[{{Heiles}(1979)}]{1979Heiles}
---. 1979, \apj, 229, 533

\bibitem[{{Heiles}(1990)}]{1990Heiles}
---. 1990, \apj, 354, 483

\bibitem[{{Houck} \& {Bregman}(1990)}]{1990Houck}
{Houck}, J.~C., \& {Bregman}, J.~N. 1990, \apj, 352, 506

\bibitem[{{Jackson} {et~al.}(2004){Jackson}, {Simon}, {Shah}, {Rathborne},
  {Heyer}, {Clemens}, \& {Bania}}]{2004Jackson}
{Jackson}, J.~M., {Simon}, R., {Shah}, R., {Rathborne}, J., {Heyer}, M.~H.,
  {Clemens}, D.~P., \& {Bania}, T.~M. 2004, in Astronomical Society of the
  Pacific Conference Series, Vol. 317, Milky Way Surveys: The Structure and
  Evolution of our Galaxy, ed. D.~{Clemens}, R.~{Shah}, \& T.~{Brainerd}, 49

\bibitem[{{Kalberla} {et~al.}(2005){Kalberla}, {Burton}, {Hartmann}, {Arnal},
  {Bajaja}, {Morras}, \& {P{\"o}ppel}}]{2005Kalberla}
{Kalberla}, P.~M.~W., {Burton}, W.~B., {Hartmann}, D., {Arnal}, E.~M.,
  {Bajaja}, E., {Morras}, R., \& {P{\"o}ppel}, W.~G.~L. 2005, \aap, 440, 775

\bibitem[{{Kalberla} {et~al.}(2007){Kalberla}, {Dedes}, {Kerp}, \&
  {Haud}}]{2007Kalberla}
{Kalberla}, P.~M.~W., {Dedes}, L., {Kerp}, J., \& {Haud}, U. 2007, \aap, 469,
  511

\bibitem[{{Kerr} \& {Lynden-Bell}(1986)}]{1986Kerr}
{Kerr}, F.~J., \& {Lynden-Bell}, D. 1986, \mnras, 221, 1023

\bibitem[{{Kerton} {et~al.}(2006){Kerton}, {Knee}, \& {Schaeffer}}]{2006Kerton}
{Kerton}, C.~R., {Knee}, L.~B.~G., \& {Schaeffer}, A.~J. 2006, \aj, 131, 1501

\bibitem[{{Koo} {et~al.}(1992){Koo}, {Heiles}, \& {Reach}}]{1992Koo}
{Koo}, B.-C., {Heiles}, C., \& {Reach}, W.~T. 1992, \apj, 390, 108

\bibitem[{{Koo} \& {McKee}(1992)}]{1992KooB}
{Koo}, B.-C., \& {McKee}, C.~F. 1992, \apj, 388, 93

\bibitem[{{Lockman}(1981)}]{1981Lockman}
{Lockman}, F.~J. 1981, \apj, 245, 459

\bibitem[{{Lockman}(1984)}]{1984Lockman}
---. 1984, \apj, 283, 90

\bibitem[{{Lockman}(2002)}]{2002Lockman}
---. 2002, \apj, 580, L47

\bibitem[{{Luna} {et~al.}(2006){Luna}, {Bronfman}, {Carrasco}, \&
  {May}}]{2006Luna}
{Luna}, A., {Bronfman}, L., {Carrasco}, L., \& {May}, J. 2006, \apj, 641, 938

\bibitem[{{Mac Low} {et~al.}(1989){Mac Low}, {McCray}, \&
  {Norman}}]{1989MacLow}
{Mac Low}, M.-M., {McCray}, R., \& {Norman}, M.~L. 1989, \apj, 337, 141

\bibitem[{{McClure-Griffiths} \& {Dickey}(2007)}]{2007McClure-Griffiths}
{McClure-Griffiths}, N.~M., \& {Dickey}, J.~M. 2007, \apj, 671, 427

\bibitem[{{McClure-Griffiths} {et~al.}(2002){McClure-Griffiths}, {Dickey},
  {Gaensler}, \& {Green}}]{2002McClure-Griffiths}
{McClure-Griffiths}, N.~M., {Dickey}, J.~M., {Gaensler}, B.~M., \& {Green},
  A.~J. 2002, \apj, 578, 176

\bibitem[{{McClure-Griffiths} {et~al.}(2005){McClure-Griffiths}, {Dickey},
  {Gaensler}, {Green}, {Haverkorn}, \& {Strasser}}]{2005McClure-Griffiths}
{McClure-Griffiths}, N.~M., {Dickey}, J.~M., {Gaensler}, B.~M., {Green}, A.~J.,
  {Haverkorn}, M., \& {Strasser}, S. 2005, \apjs, 158, 178

\bibitem[{{McClure-Griffiths} {et~al.}(2006){McClure-Griffiths}, {Ford},
  {Pisano}, {Gibson}, {Staveley-Smith}, {Calabretta}, {Dedes}, \&
  {Kalberla}}]{2006McClure-Griffiths}
{McClure-Griffiths}, N.~M., {Ford}, A., {Pisano}, D.~J., {Gibson}, B.~K.,
  {Staveley-Smith}, L., {Calabretta}, M.~R., {Dedes}, L., \& {Kalberla},
  P.~M.~W. 2006, \apj, 638, 196

\bibitem[{{Morras} {et~al.}(2000){Morras}, {Bajaja}, {Arnal}, \&
  {P{\"o}ppel}}]{2000Morras}
{Morras}, R., {Bajaja}, E., {Arnal}, E.~M., \& {P{\"o}ppel}, W.~G.~L. 2000,
  \aaps, 142, 25

\bibitem[{{Nagashima} {et~al.}(2006){Nagashima}, {Inutsuka}, \&
  {Koyama}}]{2006Nagashima}
{Nagashima}, M., {Inutsuka}, S.-i., \& {Koyama}, H. 2006, \apjl, 652, L41

\bibitem[{{Norman} \& {Ikeuchi}(1989)}]{1989Norman}
{Norman}, C.~A., \& {Ikeuchi}, S. 1989, \apj, 345, 372

\bibitem[{{Paladini} {et~al.}(2004){Paladini}, {Davies}, \&
  {DeZotti}}]{2004Paladini}
{Paladini}, R., {Davies}, R.~D., \& {DeZotti}, G. 2004, \mnras, 347, 237

\bibitem[{{Press} {et~al.}(1992){Press}, {Teukolsky}, {Vetterling}, \&
  {Flannery}}]{1992Press}
{Press}, W.~H., {Teukolsky}, S.~A., {Vetterling}, W.~T., \& {Flannery}, B.~P.
  1992, {Numerical recipes in C. The art of scientific computing} (Cambridge:
  University Press, |c1992, 2nd ed.)

\bibitem[{{Putman} {et~al.}(2003){Putman}, {Bland-Hawthorn}, {Veilleux},
  {Gibson}, {Freeman}, \& {Maloney}}]{2003PutmanB}
{Putman}, M.~E., {Bland-Hawthorn}, J., {Veilleux}, S., {Gibson}, B.~K.,
  {Freeman}, K.~C., \& {Maloney}, P.~R. 2003, \apj, 597, 948

\bibitem[{{Putman} {et~al.}(2002){Putman}, {de Heij}, {Staveley-Smith},
  {Braun}, {Freeman}, {Gibson}, {Burton}, {Barnes}, {Banks}, {Bhathal}, {de
  Blok}, {Boyce}, {Disney}, {Drinkwater}, {Ekers}, {Henning}, {Jerjen},
  {Kilborn}, {Knezek}, {Koribalski}, {Malin}, {Marquarding}, {Minchin},
  {Mould}, {Oosterloo}, {Price}, {Ryder}, {Sadler}, {Stewart}, {Stootman},
  {Webster}, \& {Wright}}]{2002Putman}
{Putman}, M.~E. {et~al.} 2002, \aj, 123, 873

\bibitem[{{Richter} {et~al.}(2001{\natexlab{a}}){Richter}, {Savage}, {Wakker},
  {Sembach}, \& {Kalberla}}]{2001RichterB}
{Richter}, P., {Savage}, B.~D., {Wakker}, B.~P., {Sembach}, K.~R., \&
  {Kalberla}, P.~M.~W. 2001{\natexlab{a}}, \apj, 549, 281

\bibitem[{{Richter} {et~al.}(2001{\natexlab{b}}){Richter}, {Sembach}, {Wakker},
  {Savage}, {Tripp}, {Murphy}, {Kalberla}, \& {Jenkins}}]{2001RichterA}
{Richter}, P., {Sembach}, K.~R., {Wakker}, B.~P., {Savage}, B.~D., {Tripp},
  T.~M., {Murphy}, E.~M., {Kalberla}, P.~M.~W., \& {Jenkins}, E.~B.
  2001{\natexlab{b}}, \apj, 559, 318

\bibitem[{{Rougoor} \& {Oort}(1960)}]{1960Rougoor}
{Rougoor}, G.~W., \& {Oort}, J.~H. 1960, Proceedings of the National Academy of
  Science, 46, 1

\bibitem[{{Sato} {et~al.}(2007)}]{2007Sato}
{Sato}, M. {et~al.} 2007, \pasj, 59, 743

\bibitem[{{Shapiro} \& {Field}(1976)}]{1976Shapiro}
{Shapiro}, P.~R., \& {Field}, G.~B. 1976, \apj, 205, 762

\bibitem[{{Stanimirovi{\'c}} {et~al.}(2006){Stanimirovi{\'c}}, {Putman},
  {Heiles}, {Peek}, {Goldsmith}, {Koo}, {Kr{\v c}o}, {Lee}, {Mock}, {Muller},
  {Pandian}, {Parsons}, {Tang}, \& {Werthimer}}]{2006Stanimirovic}
{Stanimirovi{\'c}}, S. {et~al.} 2006, \apj, 653, 1210

\bibitem[{{Staveley-Smith} {et~al.}(1996){Staveley-Smith}, {Wilson}, {Bird},
  {Disney}, {Ekers}, {Freeman}, {Haynes}, {Sinclair}, {Vaile}, {Webster}, \&
  {Wright}}]{1996StaveleySmith}
{Staveley-Smith}, L. {et~al.} 1996, Publications of the Astronomical
  Society of Australia, 13, 243

\bibitem[{{Stil} {et~al.}(2006){Stil}, {Lockman}, {Taylor}, {Dickey}, {Kavars},
  {Martin}, {Rothwell}, {Boothroyd}, \& {McClure-Griffiths}}]{2006Stil}
{Stil}, J.~M. {et~al.} 2006, \apj, 637, 366

\bibitem[{{Tomisaka} \& {Ikeuchi}(1988)}]{1988Tomisaka}
{Tomisaka}, K., \& {Ikeuchi}, S. 1988, \apj, 330, 695

\bibitem[{{Vall{\'e}e}(2008)}]{2008Vallee}
{Vall{\'e}e}, J.~P. 2008, \aj, 135, 1301

\bibitem[{{van Woerden} {et~al.}(1957){van Woerden}, {Rougoor}, \&
  {Oort}}]{1957vanWoerden}
{van Woerden}, H., {Rougoor}, W., \& {Oort}, J. 1957, {Comptes rendus de
  l'Acad\'{e}mie des sciences}, 244, 1691

\bibitem[{{Wakker}(1991)}]{1991bWakker}
{Wakker}, B.~P. 1991, \aap, 250, 499

\bibitem[{{Wakker} \& {van Woerden}(1991)}]{1991Wakker}
{Wakker}, B.~P., \& {van Woerden}, H. 1991, \aap, 250, 509

\bibitem[{{Wakker} {et~al.}(2008){Wakker}, {York}, {Wilhelm}, {Barentine},
  {Richter}, {Beers}, {Ivezi{\'c}}, \& {Howk}}]{2008Wakker}
{Wakker}, B.~P., {York}, D.~G., {Wilhelm}, R., {Barentine}, J.~C., {Richter},
  P., {Beers}, T.~C., {Ivezi{\'c}}, {\v Z}., \& {Howk}, J.~C. 2008, \apj, 672,
  298

\bibitem[{{Weiner} {et~al.}(2001){Weiner}, {Vogel}, \& {Williams}}]{2001Weiner}
{Weiner}, B.~J., {Vogel}, S.~N., \& {Williams}, T.~B. 2001, in Astronomical
  Society of the Pacific Conference Series, Vol. 240, Gas and Galaxy Evolution,
  ed. J.~E. {Hibbard}, M.~{Rupen}, \& J.~H. {van Gorkom}, 515

\bibitem[{{Williams}(1973)}]{1973Williams}
{Williams}, D.~R.~W. 1973, \aaps, 8, 505

\end{thebibliography}

\LongTables

\begin{deluxetable}{rrrrrrccr}
\tabletypesize{\footnotesize}
\tablewidth{0pt}
\tablecaption{HI Clouds - Observed Properties\label{catalog}} 
\tablehead{\colhead{$l$}&\colhead{$b$}&\colhead{$V_{\mathrm{LSR}}$}&\colhead{$T_{\mathrm{pk}}$\tablenotemark{a}}&\colhead{$\Delta v$}&\colhead{$N_{HI}$}&\colhead{$\theta_{\mathrm{min}}\times\theta_{\mathrm{maj}}$\tablenotemark{b}}&\colhead{$M_{HI}d^{-2}$\,\,\,\tablenotemark{c}}&\colhead{Notes}\\ 
\colhead{(deg)}&\colhead{(deg)}&\colhead{(km~s$^{-1}$)}&\colhead{(K)}&\colhead{(km~s$^{-1}$)}&\colhead{($\times
  10^{19}$~cm$^{-2}$)}&\colhead{(\,\,$\arcmin$\, $\times$\, $\arcmin$\,\,)}&\colhead{($M_{\odot}$ kpc$^{-2}$)}} 
\startdata
$324.78$ & $13.39$ & $-91.1 \pm 4.2$ & $0.20$ & $8.8 \pm 3.3$ & $0.34 \pm 0.18$ & $24 \times 38$ & $6.3 $ & \\ 
$324.87$ & $-8.30$ & $-73.2 \pm 2.2$ & $1.16$ & $12.7 \pm 1.1$ & $2.87 \pm 0.31$ & $17 \times 36$ & $18.6 $ & \\ 
$325.11$ & $-8.56$ & $-80.5 \pm 6.3$ & $0.34$ & $18.2 \pm 3.9$ & $1.20 \pm 0.37$ & $28 \times 36$ & $24.0 $ & \\ 
$325.34$ & $-4.61$ & $-94.2 \pm 1.8$ & $0.44$ & $6.1 \pm 1.3$ & $0.52 \pm 0.14$ & $27 \times 29$ & $20.5 $ & \\ 
$325.38$ & $5.14$ & $-88.4 \pm 2.2$ & $0.78$ & $10.3 \pm 1.2$ & $1.56 \pm 0.24$ & $26 \times 40$ & $29.3 $ & \\ 
\enddata
\tablecomments{Observed properties of \HI\ clouds in the GASS pilot
  region. The complete version of Table \ref{catalog} is available in
  the electronic edition of the Astrophysical Journal. We provide this
  sample as a guide to its content. Descriptions of each property are
  presented in \S \ref{observedcolumns}. Clouds that have been
  cataloged elsewhere in the literature are noted by the following labels:\\ 
  1: detected by \citet{2002Putman} and\\
  2: detected by \citet{1991Wakker}.\\
  Although the clouds detected elsewhere do not necessarily have the
  exact Galactic coordinates and $V_{\mathrm{LSR}}$ as listed here,
  it is likely that they are the same cloud and that the
  differences are a result of observational constraints. Also, we note
  that \citet{2000Morras} detected \HI\ in some areas of these
  clouds but such detections were not identified as individual
  objects.} 
\tablenotetext{a}{Uncertainties in $T_{\mathrm{pk}}$ are $0.07$~K.}
\tablenotetext{b}{Uncertainties in the maximum angular extents are
  dominated by background levels surrounding the cloud and are assumed
  to be $25\%$ of the estimated values.} 
\tablenotetext{c}{Mass uncertainties are dominated by the interactive process
  used in mass determination and are assumed to be $40\%$ of the
  estimated values.}
\end{deluxetable}

\begin{deluxetable}{rrrrrrrrr}
\tabletypesize{\footnotesize}
\tablewidth{0pt}
\tablecaption{Tangent Point \HI\ Clouds - Derived Properties \label{catalogtangent}}
\tablehead{
\colhead{$l$}&\colhead{$b$}&\colhead{$V_{\mathrm{LSR}}$}&\colhead{$V_{\mathrm{dev}}$}&\colhead{$d$}&\colhead{$R$\,\,\tablenotemark{a}}&\colhead{$z$}&\colhead{$r$}&\colhead{$M_{HI}$}\\
\colhead{(deg)}&\colhead{(deg)}&\colhead{(km~s$^{-1}$)}&\colhead{(km~s$^{-1}$)}&\colhead{(kpc)}&\colhead{(kpc)}&\colhead{(kpc)}&\colhead{(pc)}&\colhead{($M_{\odot}$)}}
\startdata
$325.34$ & $-4.61$ & $-94.2\pm 1.8$ & $ -1.7 \pm 3.5$ & $ 7.0 \pm 1.5$ & $ 4.8^{+0.4}$ & $ -0.56 \pm 0.12$ & $ 29 \pm 8$ & $ 1000 \pm 600$ \\ 
$325.69$ & $-2.87$ & $-98.8\pm 0.7$ & $ -3.9 \pm 3.1$ & $ 7.0 \pm 1.5$ & $ 4.8^{+0.4}$ & $ -0.35 \pm 0.07$ & $ 25 \pm 7$ & $ 470 \pm 270$ \\ 
$325.95$ & $-3.33$ & $-103.2\pm 0.9$ & $ -6.1 \pm 3.1$ & $ 7.0 \pm 1.4$ & $ 4.8^{+0.4}$ & $ -0.41 \pm 0.08$ & $ 34 \pm 9$ & $ 700 \pm 400$ \\ 
$326.34$ & $-4.47$ & $-118.0\pm 3.6$ & $ -19.7 \pm 4.7$ & $ 7.1 \pm 1.3$ & $ 4.7^{+0.3}$ & $ -0.55 \pm 0.10$ & $ 27 \pm 7$ & $ 270 \pm 140$ \\ 
$327.65$ & $13.25$ & $-108.3\pm 6.6$ & $ -4.5 \pm 7.3$ & $ 7.4 \pm 1.9$ & $ 4.5^{+0.6}$ & $ 1.69 \pm 0.44$ & $ 38 \pm 12$ & $ 600 \pm 400$ \\ 
$328.24$ & $2.67$ & $-135.0\pm 2.5$ & $ -24.7 \pm 3.9$ & $ 7.2 \pm 1.4$ & $ 4.5^{+0.4}$ & $ 0.34 \pm 0.06$ & $ 24 \pm 6$ & $ 1200 \pm 700$ \\ 
$329.79$ & $3.74$ & $-112.0\pm 1.5$ & $ -0.8 \pm 3.4$ & $ 7.4 \pm 1.9$ & $ 4.3^{+0.6}$ & $ 0.48 \pm 0.12$ & $ 30 \pm 9$ & $ 1800 \pm 1200$ \\ 
$330.12$ & $-3.40$ & $-129.3\pm 2.0$ & $ -18.8 \pm 3.6$ & $ 7.4 \pm 0.8$ & $ 4.2^{+0.2}$ & $ -0.44 \pm 0.05$ & $ 30 \pm 6$ & $ 120 \pm 50$ \\ 
$330.79$ & $4.34$ & $-114.1\pm 1.9$ & $ -5.5 \pm 3.6$ & $ 7.4 \pm 0.9$ & $ 4.2^{+0.3}$ & $ 0.56 \pm 0.07$ & $ 24 \pm 5$ & $ 1100 \pm 500$ \\ 
$330.85$ & $13.32$ & $-138.3\pm 10.1$ & $ -31.6 \pm 10.6$ & $ 7.6 \pm 0.6$ & $ 4.1^{+0.1}$ & $ 1.76 \pm 0.14$ & $ 36 \pm 7$ & $ 240 \pm 100$ \\ 
$330.90$ & $6.35$ & $-120.3\pm 7.8$ & $ -11.1 \pm 8.4$ & $ 7.5 \pm 0.9$ & $ 4.1^{+0.2}$ & $ 0.83 \pm 0.10$ & $ 25 \pm 5$ & $ 440 \pm 200$ \\ 
$330.91$ & $5.74$ & $-115.0\pm 4.5$ & $ -5.7 \pm 5.4$ & $ 7.5 \pm 0.9$ & $ 4.1^{+0.3}$ & $ 0.75 \pm 0.09$ & $ 30 \pm 7$ & $ 110 \pm 50$ \\ 
$331.04$ & $12.43$ & $-114.6\pm 6.3$ & $ -6.5 \pm 7.0$ & $ 7.6 \pm 0.9$ & $ 4.1^{+0.3}$ & $ 1.64 \pm 0.20$ & $ 48 \pm 10$ & $ 1000 \pm 500$ \\ 
$331.06$ & $3.87$ & $-119.7\pm 2.1$ & $ -9.2 \pm 3.6$ & $ 7.5 \pm 0.9$ & $ 4.1^{+0.3}$ & $ 0.50 \pm 0.06$ & $ 27 \pm 6$ & $ 2400 \pm 1100$ \\ 
$331.20$ & $10.53$ & $-109.0\pm 4.5$ & $ -1.3 \pm 5.4$ & $ 7.6 \pm 1.0$ & $ 4.1^{+0.3}$ & $ 1.38 \pm 0.18$ & $ 26 \pm 6$ & $ 280 \pm 140$ \\ 
$331.25$ & $-5.41$ & $-113.2\pm 3.3$ & $ -4.5 \pm 4.4$ & $ 7.5 \pm 1.0$ & $ 4.1^{+0.3}$ & $ -0.71 \pm 0.09$ & $ 48 \pm 10$ & $ 5700 \pm 2700$ \\ 
$331.33$ & $-3.40$ & $-135.5\pm 3.0$ & $ -26.8 \pm 4.2$ & $ 7.5 \pm 0.7$ & $ 4.1^{+0.2}$ & $ -0.44 \pm 0.04$ & $ 46 \pm 9$ & $ 300 \pm 130$ \\ 
$331.46$ & $3.40$ & $-112.4\pm 3.6$ & $ -3.4 \pm 4.7$ & $ 7.5 \pm 1.0$ & $ 4.1^{+0.3}$ & $ 0.44 \pm 0.06$ & $ 41 \pm 9$ & $ 4900 \pm 2300$ \\ 
$332.64$ & $8.16$ & $-115.0\pm 2.4$ & $ -4.9 \pm 3.9$ & $ 7.6 \pm 1.1$ & $ 3.9^{+0.3}$ & $ 1.08 \pm 0.15$ & $ 28 \pm 6$ & $ 120 \pm 60$ \\ 
$332.66$ & $-5.34$ & $-115.7\pm 1.1$ & $ -4.6 \pm 3.2$ & $ 7.6 \pm 1.1$ & $ 3.9^{+0.3}$ & $ -0.71 \pm 0.10$ & $ 31 \pm 7$ & $ 2000 \pm 1000$ \\ 
$332.73$ & $-3.47$ & $-112.7\pm 2.0$ & $ -0.7 \pm 3.6$ & $ 7.6 \pm 1.1$ & $ 3.9^{+0.3}$ & $ -0.46 \pm 0.06$ & $ 20 \pm 5$ & $ 800 \pm 400$ \\ 
$332.79$ & $-6.41$ & $-132.2\pm 5.5$ & $ -20.4 \pm 6.2$ & $ 7.6 \pm 0.8$ & $ 3.9^{+0.2}$ & $ -0.85 \pm 0.09$ & $ 33 \pm 7$ & $ 800 \pm 400$ \\ 
$332.86$ & $-2.73$ & $-143.7\pm 2.7$ & $ -29.7 \pm 4.0$ & $ 7.6 \pm 0.8$ & $ 3.9^{+0.2}$ & $ -0.36 \pm 0.04$ & $ 25 \pm 5$ & $ 340 \pm 150$ \\ 
$332.87$ & $-2.07$ & $-151.4\pm 5.7$ & $ -37.4 \pm 6.4$ & $ 7.6 \pm 0.8$ & $ 3.9^{+0.2}$ & $ -0.27 \pm 0.03$ & $ 24 \pm 5$ & $ 230 \pm 100$ \\ 
$333.18$ & $-9.24$ & $-114.4\pm 1.5$ & $ -1.0 \pm 3.4$ & $ 7.7 \pm 1.1$ & $ 3.8^{+0.3}$ & $ -1.23 \pm 0.17$ & $ 37 \pm 8$ & $ 460 \pm 220$ \\ 
$333.19$ & $-5.61$ & $-128.8\pm 2.9$ & $ -14.6 \pm 4.2$ & $ 7.6 \pm 1.0$ & $ 3.8^{+0.2}$ & $ -0.75 \pm 0.10$ & $ 33 \pm 7$ & $ 900 \pm 400$ \\ 
$333.20$ & $-3.87$ & $-116.2\pm 2.7$ & $ -1.7 \pm 4.1$ & $ 7.6 \pm 1.1$ & $ 3.8^{+0.3}$ & $ -0.51 \pm 0.07$ & $ 30 \pm 7$ & $ 900 \pm 400$ \\ 
$333.26$ & $-5.21$ & $-115.4\pm 1.2$ & $ -1.0 \pm 3.2$ & $ 7.6 \pm 1.1$ & $ 3.8^{+0.3}$ & $ -0.69 \pm 0.10$ & $ 46 \pm 10$ & $ 5100 \pm 2500$ \\ 
$333.51$ & $-9.44$ & $-133.4\pm 8.0$ & $ -21.4 \pm 8.5$ & $ 7.7 \pm 0.8$ & $ 3.8^{+0.2}$ & $ -1.27 \pm 0.13$ & $ 29 \pm 6$ & $ 340 \pm 150$ \\ 
$333.53$ & $-2.67$ & $-126.2\pm 6.1$ & $ -12.5 \pm 6.8$ & $ 7.6 \pm 1.0$ & $ 3.8^{+0.2}$ & $ -0.35 \pm 0.05$ & $ 20 \pm 4$ & $ 270 \pm 130$ \\ 
$333.53$ & $3.27$ & $-130.4\pm 0.8$ & $ -16.8 \pm 3.1$ & $ 7.6 \pm 0.9$ & $ 3.8^{+0.2}$ & $ 0.43 \pm 0.05$ & $ 33 \pm 7$ & $ 550 \pm 260$ \\ 
$333.67$ & $-3.20$ & $-122.6\pm 4.3$ & $ -7.4 \pm 5.3$ & $ 7.6 \pm 1.0$ & $ 3.8^{+0.3}$ & $ -0.43 \pm 0.06$ & $ 23 \pm 5$ & $ 550 \pm 270$ \\ 
$333.80$ & $-4.67$ & $-132.5\pm 1.7$ & $ -16.0 \pm 3.4$ & $ 7.7 \pm 0.9$ & $ 3.8^{+0.2}$ & $ -0.62 \pm 0.07$ & $ 19 \pm 4$ & $ 280 \pm 130$ \\ 
$334.13$ & $-2.60$ & $-124.8\pm 1.8$ & $ -8.2 \pm 3.5$ & $ 7.7 \pm 1.0$ & $ 3.7^{+0.3}$ & $ -0.35 \pm 0.05$ & $ 23 \pm 5$ & $ 230 \pm 110$ \\ 
$334.33$ & $3.80$ & $-124.4\pm 2.2$ & $ -5.2 \pm 3.7$ & $ 7.7 \pm 1.0$ & $ 3.7^{+0.3}$ & $ 0.51 \pm 0.07$ & $ 29 \pm 6$ & $ 400 \pm 190$ \\ 
$334.46$ & $-4.47$ & $-123.0\pm 2.3$ & $ -1.7 \pm 3.8$ & $ 7.7 \pm 1.1$ & $ 3.7^{+0.3}$ & $ -0.60 \pm 0.08$ & $ 38 \pm 9$ & $ 2000 \pm 1000$ \\ 
$334.47$ & $3.00$ & $-127.3\pm 1.8$ & $ -5.8 \pm 3.5$ & $ 7.7 \pm 1.0$ & $ 3.7^{+0.3}$ & $ 0.40 \pm 0.05$ & $ 31 \pm 7$ & $ 900 \pm 400$ \\ 
$334.60$ & $-3.00$ & $-134.3\pm 4.4$ & $ -12.5 \pm 5.3$ & $ 7.7 \pm 1.0$ & $ 3.6^{+0.2}$ & $ -0.40 \pm 0.05$ & $ 26 \pm 6$ & $ 570 \pm 270$ \\ 
$334.80$ & $-7.82$ & $-136.3\pm 2.2$ & $ -15.0 \pm 3.7$ & $ 7.8 \pm 1.0$ & $ 3.6^{+0.2}$ & $ -1.06 \pm 0.14$ & $ 42 \pm 9$ & $ 560 \pm 270$ \\ 
$334.93$ & $-11.95$ & $-131.5\pm 8.0$ & $ -12.0 \pm 8.6$ & $ 7.9 \pm 1.0$ & $ 3.6^{+0.2}$ & $ -1.63 \pm 0.21$ & $ 35 \pm 8$ & $ 570 \pm 270$ \\ 
$335.00$ & $0.80$ & $-136.5\pm 1.9$ & $ -14.4 \pm 3.5$ & $ 7.7 \pm 1.0$ & $ 3.6^{+0.2}$ & $ 0.11 \pm 0.01$ & $ 28 \pm 6$ & $ 220 \pm 100$ \\ 
$335.01$ & $-7.22$ & $-123.6\pm 2.2$ & $ -2.5 \pm 3.7$ & $ 7.8 \pm 1.1$ & $ 3.6^{+0.3}$ & $ -0.98 \pm 0.14$ & $ 30 \pm 7$ & $ 220 \pm 110$ \\ 
$335.07$ & $-6.62$ & $-128.2\pm 1.5$ & $ -7.0 \pm 3.3$ & $ 7.8 \pm 1.1$ & $ 3.6^{+0.3}$ & $ -0.89 \pm 0.12$ & $ 25 \pm 6$ & $ 200 \pm 90$ \\ 
$335.07$ & $-4.88$ & $-123.4\pm 2.8$ & $ -1.9 \pm 4.1$ & $ 7.7 \pm 1.1$ & $ 3.6^{+0.3}$ & $ -0.66 \pm 0.09$ & $ 31 \pm 7$ & $ 610 \pm 300$ \\ 
$335.13$ & $-4.40$ & $-142.7\pm 4.8$ & $ -20.7 \pm 5.7$ & $ 7.7 \pm 0.8$ & $ 3.6^{+0.2}$ & $ -0.59 \pm 0.06$ & $ 38 \pm 8$ & $ 800 \pm 400$ \\ 
$335.13$ & $3.00$ & $-139.4\pm 2.3$ & $ -17.2 \pm 3.8$ & $ 7.7 \pm 0.9$ & $ 3.6^{+0.2}$ & $ 0.40 \pm 0.05$ & $ 39 \pm 8$ & $ 900 \pm 400$ \\ 
$335.20$ & $-5.88$ & $-124.5\pm 3.0$ & $ -1.9 \pm 4.2$ & $ 7.8 \pm 1.1$ & $ 3.6^{+0.3}$ & $ -0.79 \pm 0.11$ & $ 24 \pm 5$ & $ 180 \pm 90$ \\ 
$335.20$ & $4.54$ & $-133.5\pm 2.6$ & $ -10.7 \pm 4.0$ & $ 7.7 \pm 1.0$ & $ 3.6^{+0.3}$ & $ 0.61 \pm 0.08$ & $ 28 \pm 6$ & $ 350 \pm 170$ \\ 
$335.27$ & $-2.47$ & $-126.6\pm 2.2$ & $ -2.2 \pm 3.7$ & $ 7.7 \pm 1.1$ & $ 3.6^{+0.3}$ & $ -0.33 \pm 0.05$ & $ 32 \pm 7$ & $ 3200 \pm 1600$ \\ 
$335.27$ & $2.13$ & $-127.0\pm 2.0$ & $ -2.5 \pm 3.6$ & $ 7.7 \pm 1.1$ & $ 3.6^{+0.3}$ & $ 0.29 \pm 0.04$ & $ 25 \pm 6$ & $ 600 \pm 300$ \\ 
$335.33$ & $4.07$ & $-128.8\pm 1.5$ & $ -4.3 \pm 3.4$ & $ 7.7 \pm 1.1$ & $ 3.5^{+0.3}$ & $ 0.55 \pm 0.08$ & $ 53 \pm 12$ & $ 2300 \pm 1100$ \\ 
$335.74$ & $-5.07$ & $-151.0\pm 1.6$ & $ -22.4 \pm 3.4$ & $ 7.8 \pm 0.8$ & $ 3.5^{+0.2}$ & $ -0.69 \pm 0.07$ & $ 58 \pm 12$ & $ 3200 \pm 1400$ \\ 
$335.94$ & $-4.07$ & $-141.2\pm 2.3$ & $ -11.5 \pm 3.8$ & $ 7.8 \pm 1.0$ & $ 3.5^{+0.2}$ & $ -0.55 \pm 0.07$ & $ 32 \pm 7$ & $ 1100 \pm 500$ \\ 
$336.00$ & $2.40$ & $-132.6\pm 0.8$ & $ -3.5 \pm 3.1$ & $ 7.8 \pm 1.1$ & $ 3.5^{+0.3}$ & $ 0.33 \pm 0.05$ & $ 17 \pm 4$ & $ 70 \pm 30$ \\ 
$336.20$ & $3.13$ & $-129.8\pm 1.0$ & $ -0.2 \pm 3.2$ & $ 7.8 \pm 1.1$ & $ 3.4^{+0.3}$ & $ 0.43 \pm 0.06$ & $ 27 \pm 6$ & $ 1200 \pm 600$ \\ 
$336.87$ & $-5.01$ & $-141.3\pm 5.0$ & $ -13.2 \pm 5.8$ & $ 7.8 \pm 1.0$ & $ 3.3^{+0.2}$ & $ -0.68 \pm 0.09$ & $ 49 \pm 11$ & $ 5100 \pm 2400$ \\ 
$336.94$ & $6.08$ & $-139.6\pm 4.1$ & $ -12.3 \pm 5.1$ & $ 7.9 \pm 1.0$ & $ 3.3^{+0.2}$ & $ 0.83 \pm 0.11$ & $ 30 \pm 7$ & $ 490 \pm 240$ \\ 
$337.01$ & $-5.14$ & $-132.3\pm 3.2$ & $ -4.7 \pm 4.4$ & $ 7.9 \pm 1.1$ & $ 3.3^{+0.3}$ & $ -0.70 \pm 0.10$ & $ 44 \pm 10$ & $ 4000 \pm 1900$ \\ 
$337.27$ & $-2.94$ & $-127.5\pm 2.1$ & $ -0.3 \pm 3.7$ & $ 7.8 \pm 1.1$ & $ 3.3^{+0.3}$ & $ -0.40 \pm 0.06$ & $ 41 \pm 9$ & $ 12000 \pm 6000$ \\ 
$337.35$ & $-4.47$ & $-145.0\pm 1.3$ & $ -18.9 \pm 3.3$ & $ 7.9 \pm 0.9$ & $ 3.3^{+0.2}$ & $ -0.61 \pm 0.07$ & $ 26 \pm 5$ & $ 100 \pm 50$ \\ 
$337.55$ & $5.41$ & $-146.4\pm 3.3$ & $ -20.6 \pm 4.5$ & $ 7.9 \pm 0.8$ & $ 3.2^{+0.2}$ & $ 0.74 \pm 0.08$ & $ 39 \pm 8$ & $ 190 \pm 90$ \\ 
$337.61$ & $16.01$ & $-131.1\pm 18.9$ & $ -8.6 \pm 19.1$ & $ 8.2 \pm 1.1$ & $ 3.2^{+0.3}$ & $ 2.25 \pm 0.30$ & $ 32 \pm 7$ & $ 170 \pm 80$ \\ 
$337.87$ & $-2.47$ & $-145.4\pm 1.8$ & $ -16.6 \pm 3.5$ & $ 7.9 \pm 0.9$ & $ 3.2^{+0.3}$ & $ -0.34 \pm 0.04$ & $ 44 \pm 9$ & $ 3600 \pm 1700$ \\ 
$337.87$ & $2.60$ & $-137.4\pm 1.2$ & $ -8.6 \pm 3.2$ & $ 7.9 \pm 1.0$ & $ 3.2^{+0.3}$ & $ 0.36 \pm 0.05$ & $ 29 \pm 6$ & $ 230 \pm 110$ \\ 
$337.88$ & $-5.07$ & $-155.1\pm 5.4$ & $ -26.6 \pm 6.2$ & $ 7.9 \pm 0.8$ & $ 3.2^{+0.2}$ & $ -0.70 \pm 0.07$ & $ 52 \pm 11$ & $ 1100 \pm 500$ \\ 
$337.88$ & $-3.80$ & $-152.2\pm 2.0$ & $ -23.5 \pm 3.6$ & $ 7.9 \pm 0.8$ & $ 3.2^{+0.2}$ & $ -0.52 \pm 0.05$ & $ 31 \pm 6$ & $ 380 \pm 170$ \\ 
$337.94$ & $-3.13$ & $-142.5\pm 2.0$ & $ -13.9 \pm 3.6$ & $ 7.9 \pm 1.0$ & $ 3.2^{+0.3}$ & $ -0.43 \pm 0.05$ & $ 30 \pm 6$ & $ 1100 \pm 500$ \\ 
$337.94$ & $4.14$ & $-147.4\pm 4.5$ & $ -18.9 \pm 5.4$ & $ 7.9 \pm 0.9$ & $ 3.2^{+0.3}$ & $ 0.57 \pm 0.07$ & $ 40 \pm 8$ & $ 600 \pm 280$ \\ 
$337.99$ & $17.47$ & $-176.0\pm 4.1$ & $ -53.4 \pm 5.1$ & $ 8.3 \pm 0.4$ & $ 3.2^{+0.1}$ & $ 2.48 \pm 0.11$ & $ 34 \pm 6$ & $ 240 \pm 100$ \\ 
$338.40$ & $-2.33$ & $-152.8\pm 2.7$ & $ -25.1 \pm 4.0$ & $ 7.9 \pm 0.8$ & $ 3.1^{+0.2}$ & $ -0.32 \pm 0.03$ & $ 33 \pm 7$ & $ 800 \pm 400$ \\ 
$338.54$ & $-3.74$ & $-159.8\pm 4.1$ & $ -30.0 \pm 5.1$ & $ 7.9 \pm 0.7$ & $ 3.1^{+0.1}$ & $ -0.52 \pm 0.04$ & $ 43 \pm 8$ & $ 550 \pm 240$ \\ 
$339.08$ & $-3.13$ & $-134.2\pm 1.9$ & $ -0.4 \pm 3.6$ & $ 8.0 \pm 1.1$ & $ 3.0^{+0.3}$ & $ -0.43 \pm 0.06$ & $ 37 \pm 8$ & $ 2700 \pm 1300$ \\ 
$339.29$ & $-5.07$ & $-141.4\pm 2.3$ & $ -5.5 \pm 3.8$ & $ 8.0 \pm 1.0$ & $ 3.0^{+0.3}$ & $ -0.71 \pm 0.09$ & $ 40 \pm 9$ & $ 1000 \pm 500$ \\ 
$339.65$ & $8.36$ & $-140.0\pm 4.9$ & $ -6.3 \pm 5.8$ & $ 8.1 \pm 1.1$ & $ 3.0^{+0.3}$ & $ 1.17 \pm 0.15$ & $ 50 \pm 11$ & $ 800 \pm 400$ \\ 
$339.88$ & $2.13$ & $-153.9\pm 5.0$ & $ -18.1 \pm 10.3$ & $ 8.0 \pm 0.9$ & $ 2.9^{+0.2}$ & $ 0.30 \pm 0.03$ & $ 31 \pm 7$ & $ 900 \pm 400$ \\ 
$340.25$ & $-5.94$ & $-144.6\pm 5.4$ & $ -9.5 \pm 10.5$ & $ 8.0 \pm 1.0$ & $ 2.9^{+0.3}$ & $ -0.83 \pm 0.10$ & $ 30 \pm 6$ & $ 1000 \pm 500$ \\ 
$340.35$ & $-2.87$ & $-142.3\pm 2.5$ & $ -7.6 \pm 9.3$ & $ 8.0 \pm 1.0$ & $ 2.9^{+0.3}$ & $ -0.40 \pm 0.05$ & $ 43 \pm 9$ & $ 2100 \pm 1000$ \\ 
$340.41$ & $2.40$ & $-164.8\pm 3.0$ & $ -29.5 \pm 9.5$ & $ 8.0 \pm 0.8$ & $ 2.9^{+0.2}$ & $ 0.34 \pm 0.03$ & $ 35 \pm 7$ & $ 1300 \pm 600$ \\ 
$341.65$ & $-4.47$ & $-150.5\pm 4.9$ & $ -6.8 \pm 10.2$ & $ 8.1 \pm 1.0$ & $ 2.7^{+0.3}$ & $ -0.63 \pm 0.08$ & $ 33 \pm 7$ & $ 370 \pm 170$ \\ 
$342.68$ & $16.98$ & $-177.4\pm 12.4$ & $ -35.7 \pm 15.3$ & $ 8.5 \pm 0.6$ & $ 2.5^{+0.1}$ & $ 2.48 \pm 0.17$ & $ 27 \pm 5$ & $ 200 \pm 80$ \\ 
$342.91$ & $3.54$ & $-149.2\pm 3.9$ & $ 0.4 \pm 9.8$ & $ 8.1 \pm 0.8$ & $ 2.5^{+0.2}$ & $ 0.50 \pm 0.05$ & $ 31 \pm 6$ & $ 340 \pm 150$ \\ 
\enddata
\tablecomments{Derived properties of a subset of the clouds assumed to
  be located at tangent points (those with $V_{\mathrm{dev}}\leq
  0.8$~km~s$^{-1}$).} 
\tablenotetext{a}{Along a given line-of-sight, the smallest
  Galactocentric radius possible is at the tangent point. If the cloud
  is not located at the tangent point it must be further away from the
  center and the error on $R$ must be positive.}
\end{deluxetable}

\begin{deluxetable}{lcccc}
\tablecaption{Property Summary and Comparison to Lockman Clouds\label{comptable}}
\tablehead{ & \multicolumn{2}{c}{This Sample} & \multicolumn{2}{c}{Lockman (2002)}\\
\colhead{Parameter} & \colhead{Median} & \colhead{90\%\ Range} & \colhead{Median} & \colhead{90\%\ Range}}
\startdata
$T_{\mathrm{pk}}$ (K) & $0.6$ & $0.2 \rightarrow 2.2$ & $1.0$ &$0.4 \rightarrow 2.7$\\
$\Delta v$ (km~s$^{-1}$) & $12.8$ & $5.8 \rightarrow 26.2$ & $12.2$ & $5.4\rightarrow 26.3$\\
$N_{HI}$ ($\times 10^{19}$ cm$^{-2}$) & $1.4$ & $0.2 \rightarrow 2.2$ & $2$ &
$0.7\rightarrow 6.3$\\
$r$ (pc) & $32$ & $< 23 \rightarrow 50$ & $12$ &$<9.5\rightarrow 17.5$\\
$M_{HI}$ ($M_{\odot}$) & $630$ & $120 \rightarrow 4850 $ & $50$ &$12\rightarrow 290$\\
$|z|$ (pc) & $560$ & $320 \rightarrow 1690 $ & $940$ & $640\rightarrow 1210$ \\
$\sigma_{cc}$ (km~s$^{-1}$) & $16-22$ & & $15-20$ &\\
\enddata
\tablecomments{Median values of the observed halo cloud properties in
  this sample, where the number of clouds $n=403$ for $T_{\mathrm{pk}}$,
  $\Delta v$, and $N_{HI}$ and $n=81$ for remaining properties
  (of tangent point clouds) and the \cite{2002Lockman} sample, where
  $n=38$. Most properties have a large scatter about the median in
  both samples, as demonstrated by the $90\%$ range.}
\end{deluxetable}

\end{document}